\documentclass[useAMS,usenatbib, usegraphicx]{mn2e}
\usepackage{amssymb}
\let\ga=\gtrsim \let\la=\lesssim

\title[Quenching Feedback in Cosmological Simulations] {Quenching
Massive Galaxies with On-the-fly Feedback in Cosmological
Hydrodynamic Simulations}

\author[Gabor et al.]{
J. M. Gabor,$^{1}$\thanks{Email:jgabor@as.arizona.edu}
R. Dav\'e,$^{1}$ B. D. Oppenheimer$^{2,3}$ \& K. Finlator$^{4,5}$ \\
$^{1}$University of Arizona, 933 N. Cherry Ave, Tucson, AZ 85721\\
$^{2}$Leiden Observatory, Leiden University, PO Box 9513, 2300 RA
Leiden, the Netherlands\\ $^{3}$VENI Fellow\\ $^{4}$University of
California, Santa Barbara Physics, Santa Barbara, CA 93106\\
$^{5}$Hubble Fellow\\ }

\begin{document}



\maketitle
\label{firstpage}

 \begin{abstract}
   Massive galaxies today typically are not forming stars despite
   being surrounded by hot gaseous halos with short central cooling
   times.  This likely owes to some form of ``quenching feedback''
   such as merger-driven quasar activity or radio jets emerging from
   central black holes.  Here we implement heuristic prescriptions for
   these phenomena on-the-fly within cosmological hydrodynamic
   simulations.  We constrain them by comparing to observed luminosity
   functions and color-magnitude diagrams from SDSS.  We find that
   quenching from mergers alone does not produce a realistic red
   sequence, because $1-2$~Gyr after a merger the remnant accretes new
   fuel and star formation reignites.  In contrast, quenching by
   continuously adding thermal energy to hot gaseous halos
   quantitatively matches the red galaxy luminosity function and
   produces a reasonable red sequence.  Small discrepancies remain --
   a shallow red sequence slope suggests that our models underestimate
   metal production or retention in massive red galaxies, while a
   deficit of massive blue galaxies may reflect the fact that observed
   heating is intermittent rather than continuous.  Overall, injection
   of energy into hot halo gas appears to be a necessary and
   sufficient condition to broadly produce red and dead massive
   galaxies as observed.

\end{abstract}

\section{Introduction}

Observations show a distinct bimodality in galaxy colours
\citep{strateva01, baldry04, balogh04, bell04, weiner05, willmer06},
with massive galaxies today generally being ellipticals having little
star formation or cold gas, while less massive galaxies tend to be
blue spirals with cold gas.  Despite being recognized in the earliest
observations of galaxies, the origin of this bimodality remains poorly
understood.  The lack of galaxies in the region between the red
sequence of ellipticals and blue cloud of spirals (often called the
green valley) indicates that star formation must be quenched rapidly
in order to transform blue galaxies into red ones \citep{bell04,
  blanton06}.  The preponderance of active galactic nuclei (AGN) in
the green valley suggests that AGN are somehow connected to the
process of quenching star formation \citep[e.g. ][]{schawinski09}.
Several physical mechanisms have been proposed to effect this
transformation, with two frontrunners garnering much attention
recently: feedback associated with galaxy mergers, and quenching due
to virial shock heating of accreted gas.  Both of these invoke AGN as
a crucial energy source, namely quasars in the former case and radio
jets in the latter.

Understanding the origin of massive, passive ``red and dead'' galaxies
remains a problem for models of galaxy evolution.  Hydrodynamic
simulations of gas-rich galaxy mergers have been shown to induce
starbursts and fuel powerful quasars \citep{hernquist89, barnes92,
dimatteo05}.  Observations lend support to the idea of a connection
between mergers and powerful AGN (\citealp{hutchings88, sanders96,
canalizo01, bennert08, urrutia08}; but not all:
e.g. \citealt{bahcall97, mclure99, grogin05, gabor09, georgakakis09}).  In
merger simulations, the combination of feedback from star formation
and black hole accretion unbinds the gas from the merger remnant,
leaving a red elliptical galaxy devoid of fuel for star formation
\citep{silk98, springel05_mergers_ellipticals, hopkins08_ellipticals}.
This is now colloquially known as ``quasar mode'' feedback.
Observations support various aspects of this picture, including
high-velocity outflows from post-starburst galaxies \citep{tremonti07}
and QSOs \citep{feruglio10}, and a peak in the fraction of AGN host
galaxies with colors between the red sequence and blue cloud
(e.g. \citealt{silverman08, schawinski09}, but see also
\citealt{silverman09, xue10}).  Recent work hints that secular
processes may induce AGN \citep{cisternas11}, but it is not clear whether such AGN alone
can quench star-formation in their host galaxies.

The presence of hot X-ray emitting gaseous halos around red and dead
galaxies in groups and clusters suggests that these may also be
associated with quenching.  In this scenario, gas falling into the
halo from the intergalactic medium (IGM) collides with stationary hot
gas already in the halo.  The resulting shock heats the gas to the
virial temperature, converting the gravitational energy of infall to
thermal energy \citep[cf.][]{eggen62, silk77, rees77, white78}.
Several recent hydrodynamic simulations show that radiative cooling
dominates over this shock heating in dark matter haloes below a rough
critical mass of $\sim10^{12} M_{\sun}$, which is suggestively close
to the mass at which the color bimodality divides galaxies.  Above
this critical mass, a stable shock forms and the halo develops a
stable hot gas halo \citep{birnboim03, keres05, keres09_coldmode}.
This hot halo will shock heat newly accreting gas, thus stalling the
ultimate fuel for star formation \citep{dekel06, birnboim07}.  Dense
filaments of gas may penetrate this hot halo, avoiding the virial
heating and providing fuel for star formation, but this is predicted
to be uncommon at redshifts less than 2 \citep{dekel06, ocvirk08,
dekel09_nature} because filaments are less dense at late times.

In massive galaxy cluster haloes with peaked density distributions
\citep[roughly half of all clusters;][]{bauer05}, the hot gas appears
to be cooling rapidly via X-ray emission \citep{fabian84}, yet UV,
optical, and NIR observations indicate only trace levels of star
formation \citep{smith97, crawford99, hicks05, quillen08, donahue10}
and little cold gas \citep[e.g.][]{salome06}.  Hence some additional
heating mechanism must counteract the X-ray cooling; this is the
well-known cooling flow problem \citep{fabian94}.  Observations of
X-ray cavities thought to result from powerful radio jets have
motivated a picture where a radio AGN in the central cluster galaxy
provides this extra heating \citep{mcnamara00, fabian00, mcnamara05}.
As gas cools from the hot halo, some of it accretes onto the massive
central black hole, inducing an AGN \citep{ciotti97}.  The kinetic
power emerging via radio jets creates bubbles of typical size $\sim
10$ kpc \citep{birzan04} in the central regions of the cluster.  These
bubbles then expand and degenerate into sound waves that effectively
isotropize outflow energy and heat the hot intracluster gas on scales
$>$30 kpc \citep{fabian03, voit05, fabian06, deyoung10}.  From gas
cooling to gas heating by the AGN, this entire cycle may repeat itself
on $\sim 10^{8-9}$ yr timescales \citep{ciotti01}.  This is now referred to as
``radio mode'' feedback \citep{croton06}.

Mechanisms unrelated to supermassive black holes may provide the
additional heating in hot X-ray haloes.  Thermal conduction in
clusters provides central heating and inhibits cooling instabilities,
although it is not sufficient to prevent cooling flows in all clusters
\citep{zakamska03,parrish09, ruszkowski10}.  Infalling galaxies or
stellar clumps may dynamically heat haloes by transferring
gravitational energy to the gas \citep{dekel08, khochfar08, johansson09,
ruszkowski10_stirring, birnboim10}.  Whatever the mechanism, the key
requirement is to heat up the cores enough to prevent substantial
cooling and star formation, while allowing central cooling times less
than the Hubble time (or even $<1$ Gyr) to persist in a substantial
fraction of clusters and groups.

Debate persists about the effectiveness of these proposed quenching
mechanisms on scales smaller than a kpc, within a single galaxy.
Various simulations use different models for black hole accretion and
feedback \citep{springel05_mergers_ellipticals, booth09, debuhr09,
levine10}, and recent work raises doubts that even a powerful quasar
could significantly affect star-forming gas throughout a galaxy
\citep{debuhr09, debuhr10}.  In radio mode feedback, the observed AGN
energy output is generally sufficient to counteract the radiative
cooling in clusters or groups \citep{mcnamara06, best06, giodini10}.
How this energy is efficiently distributed over space and time to
prevent substantial cooling and star formation remains unclear, and
cosmological models generally do not include realistic AGN radio jet
physics.  Hence while supermassive black holes are likely to be
responsible for quenching, direct ab initio simulations of this
process on a cosmological scale remain beyond current reach.

Semi-analytic models (SAMs), which marry analytic prescriptions for
baryon physics to merger trees from cosmological simulations of dark
matter, have successfully incorporated quenching mechanisms to create
reasonable red sequence populations.  Among analytic prescriptions for
gas cooling and star formation, these models typically incorporate a
starburst and bulge growth during galaxy mergers.  These merger
prescriptions are not enough to make massive galaxies red, so modelers
include some suppression of cooling in hot, massive haloes, usually
via AGN feedback \citep{croton06, bower06,somerville08}.  While most
SAMs contain such prescriptions, the exact amount of quenching
feedback required and the mechanisms by which they operate differ
widely among models, possibly because these models must make many
assumptions regarding the gas dynamics which can be quite
uncertain~\citep{lu10_cooling,delucia10}.

Hydrodynamic cosmological models follow the gas dynamics
directly, but without some explicit quenching mechanisms they
generally do not yield a realistic red sequence with appropriate
luminosity or mass functions.  These models greatly reduce the number
of free parameters relative to SAMs by tracking the dynamics of gas
inflows and outflows directly, at large computational cost.  Although
hydrodynamic models directly track the dynamics of galaxy mergers and
yield hot gas in massive haloes, they still produce only constantly
growing, blue, star-forming galaxies at the massive end of the mass
function.  Recent work has begun to include black hole fueling and
feedback.  These works have focused on reproducing the black hole
mass--bulge velocity dispersion ($M-\sigma$) relation, black hole
properties, and AGN population properties, with significant successes
\citep[e.g.][]{sijacki07, degraf10, booth09, booth10}.  Typical models
involve Bondi-Hoyle accretion of surrounding gas by the central black
hole, with some assumed fraction of the accreted rest-mass converted
to heat in the surrounding gas \citep{springel05_mergers_ellipticals,
booth09}.  These prescriptions, while pioneering, are poorly
constrained and may not properly model the dominant accretion
mechanisms \citep{booth09, debuhr09}.  They also do not focus on
producing red sequence galaxies as observed, and have had only limited
success doing so.

In this work we also employ hydrodynamic simulations, but we focus on
reproducing the observed galaxy population rather than the observed
black hole and AGN population.  Our approach is therefore somewhat
different, in a sense intermediate between fully hydrodynamical
simulations and SAMs.  Here we simply ask, how is quenching related to
the evolutionary properties of galaxies and their halos?  We apply
heuristic models for quenching by adding energy in various forms to
galaxies and their surrounding gas during the evolution of the
simulation, and ask which models are successful at producing red and
dead galaxies as observed.  We do not try to explicitly grow black
holes and account for their feedback.  In \citet{gabor10}, we conducted a precursor study on
{\it post-processed} star formation histories, and concluded that both
radio mode and quasar mode feedback could in principle produce red
galaxies if all future star formation was assumed to be suppressed,
but several other attempted mechanisms such as shutting off hot mode
accretion or recycled wind mode accretion could not.  As such, we now
focus particularly on the two popular mechanisms of radio mode and
quasar mode, with the goal of determining which one drives the
quenching of massive galaxies.

Here, as in \citet{gabor10}, we compare our simulation results
to observations of the galaxy population at redshift $z\approx 0$, and
leave detailed evolutionary studies for the future.  We consider the
red sequence in color-magnitude diagrams, and the corresponding red
galaxy luminosity function.  Due to uncertainties in dust models, we
use galaxy stellar mass functions to compare our simulated blue cloud
population observations.  Our main result is that superwinds induced
by galaxy mergers cannot create a red sequence, because galaxies
continue to accrete gas from the IGM even after a merger, while our
quenching mechanism based on hot halo gas produces a successful match
to the observed red galaxy luminosity function.  This favors radio
mode feedback as the primary driver for the formation of the massive
red galaxy population.

We describe the physics of our simulations in \S\ref{sec.simulations},
including our newly implemented quenching models in
\S\ref{sec.new_physics}.  In \S\ref{sec.observations} we describe the
observational data. We compare the results of these new models to
observed color-magnitude diagrams (CMDs) and luminosity functions
(LFs) for local galaxies in \S\ref{sec.results}.  We then consider
physical consequences of our models in \S\ref{sec.physical}, and
discuss future directions in the discussion.

\section{Simulations} 
\label{sec.simulations} 

Our simulations are run with an extended version of the N-body +
smoothed particle hydrodynamics (SPH) cosmological simulation code
GADGET-2 \citep{springel05}, as described in \citet{oppenheimer08}.
Along with gas cooling and star-formation,  our version includes
galactic winds tied to the star formation and a chemodynamical model
with sources including AGB stars and Type Ia supernovae.  In the
following subsections, we describe additional modifications to the
code to implement quenching of star formation.

For star formation in cold, dense regions, we use the two-phase model
of \citet{springel03}.  This model, based on \citet{mckee77}, assumes
that any gas particle dense enough to be Jeans unstable contains a hot
medium and cold clouds in pressure equilibrium.  Stars form from the the cold clouds with a
characteristic timescale, $\tau$.  This timescale effectively sets the
efficiency of star formation, and \citet{springel03} find that
$\tau\approx 2$ Gyrs yields galaxy models that match the observed Kennicutt
relation \citep{kennicutt98}.  The three model components exchange
mass, energy, and metals via condensation and feedback from
supernovae.

Star-forming gas particles in the simulations spawn star particles
with a probability based on the star-formation rate.  The resulting
collisionless star particle takes on a mass equal to half the original
mass of a gas particle.  The average gas particle spawns up to two
star particles in this stochastic way.

Along with the feedback implicit in the star-formation model, our
simulations include an explicit stochastic model of galactic winds
\citep{oppenheimer08}.  We eject star-forming gas particles from
galaxies by giving them a velocity kick $v_w$ \citep{springel03}.  The
probability that a gas particle will enter a wind in the first place
is the mass-loading factor $\eta$ (the ratio of mass entering winds to
the mass of stars formed) times the probability for star formation.
In our simulations $v_w$ is proportional to, and $\eta$ inversely
proportional to, the galaxy circular velocity.  These scalings, which
arise in the momentum-driven wind model of \citet{murray05}, are
motivated by observations of winds in local galaxies \citep{martin05,
rupke05}.  Once a gas particle enters a wind, we decouple it
hydrodynamically until it reaches a density 0.1 times the critical
density for star formation (up to a maximum duration of 20 kpc/$v_w$).
This decoupling is intended to represent chimneys through which
material can escape easily but that cannot be reliably modeled by the
spherically-averaging SPH algorithm.

In order to study the chemical enrichment of galaxies and the IGM, our
code includes models for stellar mass loss and Type Ia supernovae,
described in \citet{oppenheimer08}.  As the stellar population
represented by a star particle ages, mass loss from AGB stars injects
metal-enriched mass to the nearest 3 gas particles based on the
mass-loss rates from \citet{bc03}.  We estimate Type Ia supernova
rates for each star particle using \citet{scannapieco05}, with a
prompt and delayed component.  Each supernova injects both energy and
metal mass into the surrounding gas.

With this treatment of galactic winds, our simulations match broad
observations of star-forming galaxies and the IGM, including the
chemical enrichment of the IGM at $z>2$ \citep{oppenheimer06,
oppenheimer08}, the galaxy mass-metallicity relation \citep{dave07,
finlator08}, and the luminosity functions of high-redshift galaxies
\citep{dave06, finlator07}.  These simulations do not, however, yield
massive red and dead galaxies, and galaxies are far too massive and
bright at low redshifts.  We require some additional mechanisms to
halt star-formation in massive galaxies -- the hydrodynamics and
feedback models we have described above are insufficient.

\subsection{Quenching Prescriptions} 
\label{sec.new_physics}

We incorporate two independent quenching models (which may be
combined) into GADGET-2.  We are deliberately agnostic about whether
supermassive black holes or some other mechanisms provide the energy
or momentum in these feedback prescriptions.  For the first, we assume
that major galaxy mergers ultimately drive the transformation of
galaxies from blue to red.  We identify mergers in the simulation, and
add a kinetic kick to the gas in the merger remnant to expel
star-forming gas.  This burst of energy mimics the effects of a
powerful quasar or starburst~\citep[e.g.][]{dimatteo05}.

In the second quenching model, we assume that the hot gas haloes
surrounding massive galaxies facilitate quenching.  We identify these
hot, massive haloes in the simulation, then continually add thermal
energy to the halo gas in order to cut off the fuel for
star-formation.  The addition of thermal energy mimics the added heat
from radio AGN~\citep[e.g.][]{mcnamara05}, unresolved gravitational
heating~\citep[e.g.][]{dekel08},
conduction~\citep[e.g.][]{jubelgas04}, or cosmic
rays~\citep[e.g.][]{guo08}.

\subsubsection{Quenching via galaxy mergers} 
\label{sec.merger}

\begin{figure}
\includegraphics[width=84mm]{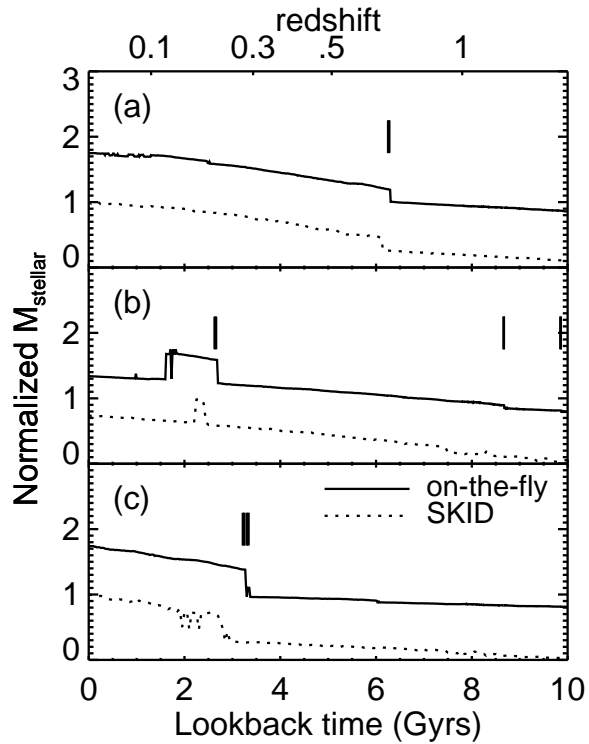}
\caption{Comparison of stellar mass histories calculated using two
  different group-finders for three galaxies (one galaxy per panel).
  Dotted lines show normalised masses identified by {\sc SKID} in
  post-processing, while solid lines show our faster on-the-fly {\sc
  FOF} masses offset by 0.75.  Vertical line segments indicate mergers
  identified on-the-fly. {\bf (a)} A galaxy with an easily identified
  major merger.  {\bf (b)} A galaxy with a major fly-by, incorrectly
  identified as a merger.  {\bf (c)} A merger where the final
  coallescence is $\sim1$ Gyr later with {\sc SKID} than with our
  on-the-fly method.  The latter two problems are sufficiently rare to
  have little impact on our results.}
\label{fig.mass_hist}
\end{figure}

Our first quenching prescription ties truncation of star formation to
galaxy mergers.  This requires identifying galaxy mergers within the
simulations on the fly, then applying some quenching process to the
particles in those mergers.  We implement empirically-motived,
high-velociy winds that eject cold gas from merger remnants.

Merger identification is based on our on-the-fly galaxy finder,
developed in \citet{oppenheimer08}.  This algorithm uses a parallel
friends-of-friends ({\sc FOF}) algorithm to group together particles
into galaxies.  We set the linking length between particles to $0.06
r_s (H(z)/H_0)^{1/3}$, where $r_s$ is the mean interparticle
separation, $H_0$ is the hubble constant ($70$ km s$^{-1}$), and
$H(z)$ is the hubble parameter as a function of redshift.
\citet{oppenheimer08} showed that stellar masses of groups found with
this prescription successfully matched those obtained in
post-processing with the slower but more physically-motivated Spline
Kernel Interpolative Denmax ({\sc SKID})
\footnote{http://www-hpcc.astro.washington.edu/tools/skid.html}
algorithm~\citep{keres05}.  To optimize for speed, we divide the
simulation volume into grid cells and only calculate distances between
particles in neighboring grid cells.  We identify groups independently
for each central processing unit, then merge groups split onto
adjacent CPUs by checking whether groups separated by less than 30
linking lengths indeed have particles that should be linked together.
The frequency with which we run the galaxy finder varies from a few
Myr to a few 10's of Myr, depending on the adaptive time-step
intervals.

We identify mergers on-the-fly by looking for persistent, rapid
increases in stellar mass growth.  This requires knowing the recent
mass history of each galaxy.  We track the 6 most recent galaxy
catalogs constructed by the galaxy finder, then connect galaxies in
the most recent catalog to their earlier most-massive progenitors.  By
tracking both the current and previous galaxy host of each particle,
we can easily identify the progenitors of a given galaxy by searching
through its constituent particles and noting their previous host
galaxies.

For a given galaxy, we identify a merger by comparing its current
stellar mass to that of its recent progenitor galaxy.  In detail, we
compare the median mass of the three most recent progenitors
($M_{\rmn{recent}}$) to the median mass of the three older progenitors
($M_{\rm older}$).  If $M_{\rm recent} / M_{\rm older} > (1+1/r)$ then
we flag the galaxy as a merger.  We use the median approach, rather
than just comparing the most recent galaxy mass to the previous one,
to avoid declaring mergers of brief encounters and fly-by's.  Since
the usual value of $r=3$ yielded good matches to the galaxy luminosity
function in \citet{gabor10}, we use the same choice here (see \S \ref{sec.appendix} for further discussion).

This method of identifying mergers likely declares the merger before
galaxies have fully merged.  Since we use all particle types in the
galaxy finder, two merging galaxies will appear as one once the
central regions of their dark matter haloes overlap.  Similarly, when
two galaxies closely pass by each other, our method will declare a
merger whether the galaxies eventually coalesce or not.  Based on the
work of \citet{maller06}, this appears to be a generic problem with
merger identification in cosmological simulations.  We mitigate this
effect partially by taking the median masses over several time steps
as described above -- rapid fly-by's will not alter the median mass
enough to cause a merger identification, although some longer, close
fly-by's do (as we show below).  In cases where the two galaxies do
finally coalesce, the precise timing of the quenching does not matter
much, since we effectively get rid of all cold gas in the remnant (see
below).  Early quenching could, however, stunt the star formation that
might naturally occur during the merging process, leading to slightly
smaller stellar masses and lower metallicities in merger remnants.  In
the end, our results are not expected to be sensitive to these
effects.

To illustrate the successes and challenges of our merger
identification method, in Figure \ref{fig.mass_hist} we compare the
stellar mass histories of galaxies identified on-the-fly to the masses
of the same galaxies identified with {\sc SKID} in post-processing.
The mass growth histories from the on-the-fly {\sc FOF} method broadly
match those from {\sc SKID}, and $\sim$60\% of galaxies with $M_{\rm
stellar}>10^{10}M_{\sun}$ in this simulation undergo at least one
merger in the last 10 Gyr.  Panel (a) shows a well-matched single
major merger.  Panel (b) illustrates the difficulty with fly-by's that
do not end up merging.  Both {\sc FOF} and {\sc SKID} combine the two
galaxies into a single galaxy at some point, and then separate them
again as the two galaxies move apart.  Among galaxies with at least
one merger, our method mis-identifies fly-by's as major mergers in
$<10$\% of the cases.  Panel (c) shows a case where detailed structure
in the mass history identified by {\sc SKID} is missed by {\sc FOF}.
In the {\sc FOF} case, it looks like a clean-cut major merger.  With
{\sc SKID}, however, the stellar mass jumps, then after a few hundred
Myrs begins to fluctuate up and down before settling in a more massive
state.  This suggests an initial fly-by, then a few more close
passages, then a final coallescence just over 1 Gyr after the initial
merger identification.  Among galaxies with at least one merger, our
method identifies one of those mergers too early by $>500$Myr in $\sim
25$\%.  In future work we may attempt to identify the final
coallescence, and avoid mis-identifying fly-by's, by invoking a delay
time of a few hundred Myr before actually declaring the merger.  Given
the broad nature of this study, such details of merger identification
do not substantially affect our results.

Once we have identified a merger, we apply energetic ejective feedback
to the star-forming gas within the merger remnant, mimicking a
powerful wind.  Observations of post-starburst galaxies and QSOs
suggest that such winds with velocities $\sim 1000$ km s$^{-1}$ or
more may be responsible for quenching star formation
\citep{tremonti07, feruglio10}.  We pursue the extreme case where all
the star-forming gas is ejected in a wind because this is probably
necessary to produce a red galaxy remnant \citep[cf. ][]{gabor10}.
Each gas particle in the merger remnant galaxy is given a velocity
kick of $v_{\rmn{kick}}$ in the direction $\vec{v} \times \vec{a}$,
where $\vec{v}$ is the pre-kick velocity vector, and $\vec{a}$ is the
acceleration vector.


\citet{tremonti07} compared post-starburst galaxies with winds
$\sim$1000 km s$^{-1}$ to low-ionization broad absorption line quasars
\citep{trump06} with winds up to $\sim 10^4$ km s$^{-1}$.  We choose
$v_{\rmn{kick}}=1500$ km s$^{-1}$ as a compromise for large-scale
outflows that have likely exited the galaxy.  Simulations with
$v_{\rmn{kick}}=1000$ km s$^{-1}$ showed little difference in the
results, since this is still well above the escape velocity for most
haloes.  As with our winds associated with star-formation, we decouple
the wind particles from hydrodynamic processes to enable escape from
the ISM, just as we do with star formation-driven winds.

As an addition to ejective feedback, we optionally implemented a
thermal feedback mechanism associated with galaxy mergers.  In this
case, we heat all the star-forming gas particles to the virial
temperature of the halo, as estimated below in \S\ref{sec.hotquench}.
It turns out that in combination with the ejective feedback, it makes
little difference to the results, since ejected gas almost never
returns to the galaxy whether we heat it or not.


\subsubsection{Quenching in hot haloes} 
\label{sec.hotquench}

\begin{figure}
\includegraphics[width=84mm]{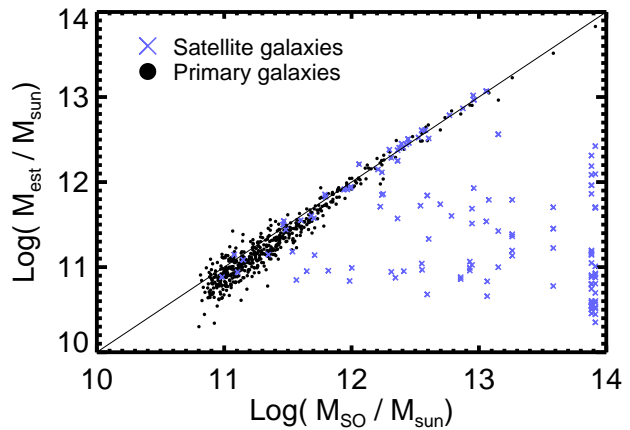}
\caption{Comparison of halo mass estimates from our on-the-fly
algorithm ($y$-axis) and a post-processing spherical overdensity
algorithm ($x$-axis).  The dotted line shows a perfect 1-to-1
correspondence.  In the on-the-fly case, we multiply the baryonic (gas
and stellar) mass found by our {\sc FOF} galaxy finder by an empirical
factor of 18 to get $M_{\rmn{est}}$ with the best fit.  This factor is
larger than the expected cosmic factor ($\sim 6$) because we tune the
{\sc FOF} linking length to match the stellar mass, not the total
baryonic mass, of galaxies.  Much of the baryonic mass remains in an
extended halo not captured by the {\sc FOF} method.  The on-the-fly
algorithm underestimates the halo masses of many satellite galaxies
(blue crosses).  This is because the spherical overdensity method
associates satellite galaxies with their massive parent halo, whereas
our {\sc FOF} galaxy finder associates them with their smaller
sub-haloes.}
\label{fig.fof_halomass}
\end{figure}

\begin{figure*}
\includegraphics[width=7in]{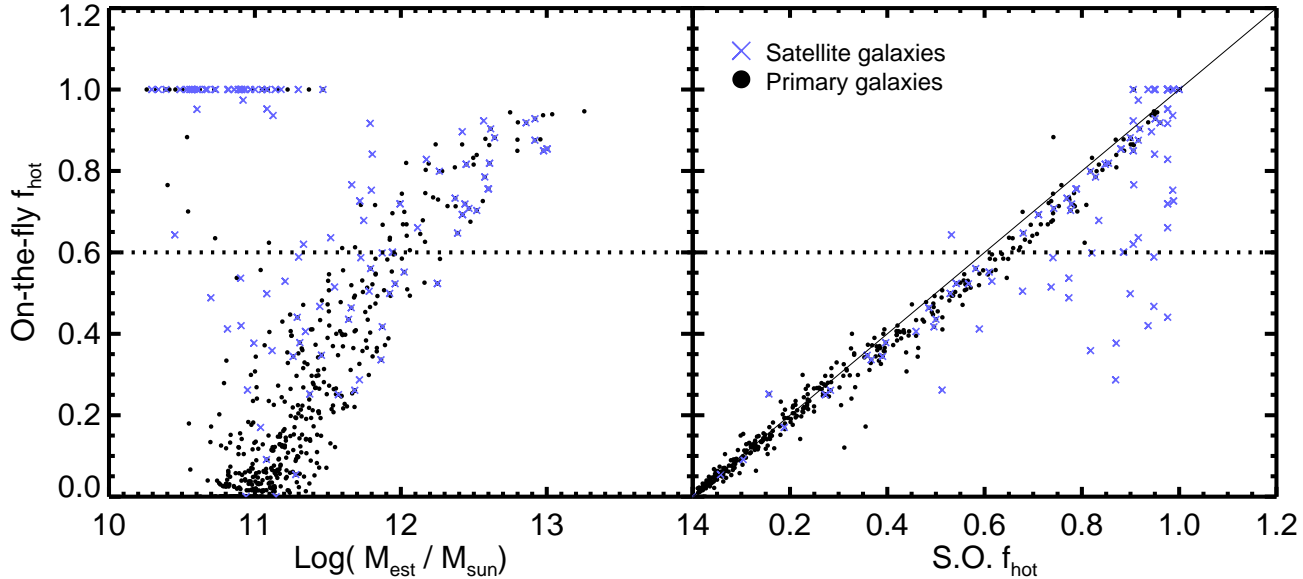}
\caption{Hot gas fractions calculated on-the-fly vs. halo mass
  ($M_{\rmn est}$) and spherical overdensity (SO) hot fraction for
  galaxies in a small volume simulation.  Black dots show central
  galaxies and blue crosses satellite galaxies (as found using SO).
  Hot gas fractions are defined as the fraction of gas within the
  virial radius above 250,000 K.
  The left panel shows that low-mass haloes have little hot gas in
  their surroundings, while increasingly larger haloes have
  increasingly high hot fractions \citep[cf. ][]{keres05}.  The
  galaxies in low-mass haloes with $f_{\rmn{hot}}=1$ are satellite
  galaxies embedded in massive, hot haloes.  The right panel compares
  our on-the-fly estimate to our post-processing, spherical
  overdensity estimate of $f_{\rmn{hot}}$, and a thin solid line shows
  a perfect 1-to-1 correspondence.  We can reliably estimate the hot
  fractions of galaxies on-the-fly in the simulation.}
\label{fig.fof_hotfrac}
\end{figure*}

The presence of a hot gaseous halo around most massive red and dead
ellipticals hints at some role in quenching.  Hence for our second
main quenching mechanism, we heat gas in the hot gaseous haloes around
galaxies.  For each galaxy identified on-the-fly, we estimate the hot
gas fraction in its halo.  In haloes dominated by hot gas, we add heat
by raising the temperature of gas particles around the galaxy.

To obtain the hot gas fractions of galactic haloes, we estimate the
virial radius corresponding to each halo, find all the gas particles
within that radius, then count up the mass on either side of a
temperature threshold defining ``hot.''  We estimate the halo virial
radius, $R_{\rmn{vir}},$ using the equation $\rho_{\rmn{vir}} =
M_{\rmn{est}} / (4/3 \pi R_{\rmn{vir}}^3)$. Here, $\rho_{\rmn{vir}}$
is the virial density for collapsed structures at the appropriate
epoch, and $M_{\rmn{est}}$ is an estimated halo mass.  We find
$M_{\rmn{est}}$ by scaling the baryonic mass of the galaxy from the
group finder by a factor of 18.  As shown in Figure
\ref{fig.fof_halomass}, this empirical scale factor value best
reproduces the total halo masses found using a more thorough spherical
overdensity algorithm in post-processing.  This departs significantly
from the cosmic value of $\Omega_M / \Omega_{\rmn{baryon}} \simeq 6$
because our on-the-fly group finder purposely does not capture the
full radial extent of the halo.  Our {\sc FOF} linking length is
chosen to match galaxy stellar masses, so it is too small to capture
much of the outer halo.  Nevertheless, the empirical factor of 18
nicely estimates tht total halo mass.

We add the mass of each gas particle within the virial radius to a
tally of the hot gas mass if the particle's temperature exceeds
250,000 K; otherwise, we consider it ``cold'' or warm for the purpose
of identifying hot gas haloes.  We choose 250,000 (or $10^{5.4}$) K
because this separates the main bulk of cooler IGM that eventually
cools into galaxies from the hot tail of the IGM from which cooling
and forming stars is rare \citep[cf. ][]{keres05,
keres09_coldmode,gabor10}.

Once we have measured the hot and cold gas masses within the virial
radius of each galaxy, we apply the quenching heat if $f_{\rmn{hot}}
\equiv m_{\rmn{hot}} / (m_{\rmn{hot}} + m_{\rmn{cold}}) > 0.6$.  A hot
fraction of 0.6 corresponds roughly to halo masses of $10^{12}
M_{\sun}$ \citep{keres05}.  In \citet{gabor10}, we found that
quenching star formation in haloes with of $M>10^{12}M_{\sun}$ or
$f_{\rmn{hot}} > 0.6$ led to luminosity functions that best reproduce
those observed.

Figure \ref{fig.fof_hotfrac} illustrates the effectiveness of our hot
fraction estimate in a small simulation.  Our on-the-fly hot fraction
calculation accurately reproduces the trend noted by \citet{keres05}:
halos above $\sim 10^{12} M_{\rmn{\sun}}$ have hot fractions above
$\sim 0.6$ (left panel), and the transition between halos dominated by
cold and hot gas is fairly abrupt.  Outliers from the trend with low
masses and $f_{\rmn{hot}}=1$ are small satellite galaxies embedded in
the hot halo of a larger galaxy.  In the right panel, we compare our
on-the-fly hot fraction estimate to that from our post-processing,
spherical overdensity method.  With a high degree of reliability, we
can determine whether the gaseous halo surrounding a galaxy is hot.

In {\sc FOF} galaxies above the hot fraction threshold, we raise gas
particles to the estimated virial temperature of the halo.  In our
base model, we only raise the temperatures of particles that are {\it
not} star-forming, i.e. that are not within the existing cold ISM of
galaxies within the halo.  We also do not raise the temperature of
particles already exceeding the virial temperature.  We use
equation~59 from \citet{voit05_review} to estimate the virial
temperature, using $M_{\rmn{est}}$ for the halo virial mass:
\begin{equation} \label{eq:T_M}
 k_B T = (8.2 \rmn{keV}) \left( \frac{M_{\rmn{est}}}{10^{15} h^{-1}M_{\sun}} \frac{H(z)}{H_0}\right)^{2/3}.
\end{equation}
Here, $k_B$ is Boltzmann's constant, $H(z)$ is the redshift-dependent
Hubble parameter, and $H_0$ is the Hubble constant (at $z=0$).  In
practice, we found that instantaneously heating particles to the
virial temperature sometimes led to unphysically high hydrodynamic
accelerations.  To prevent adaptive time-steps from becoming too
short, we instead double the entropy of the gas particle until it
reaches the virial temperature.  Through this heating mechanism, we
prevent gas from ever cooling and forming stars.

A key aspect of our mechanism is that energy is continually added to
keep particles hot.  Gas within the hot halo that cools below
$T_{\rm vir}$, or accreted gas that is below $T_{\rm vir}$, is heated
as described above during every time step of the simulation.  This is
perhaps overkill; to suppress star formation, it may not be necessary
to keep gas at or above the virial temperature, nor to do so
at every timestep.  Here, however, we will make this maximal heating
assumption to ensure star formation is suppressed, even though we will
discuss later that the energetics of this are quite extreme.  We will
explore relaxing the heating assumption in future work.

Given the overall success of our hot gas quenching model, we explored
many variants on our base model:

\begin{enumerate}
\item We heat {\it all} gas within halos to the virial tamperature,
including star-forming gas within the ISM of galaxies.

\item We use a criterion based directly on the halo mass (rather than
hot gas fraction).  While there is a close correlation between hot gas
haloes and massive dark matter haloes (cf. Figure
\ref{fig.fof_hotfrac}, \citealt{birnboim03,keres05}), some models have
had success quenching star formation above a particular halo mass,
typically around $10^{12}M_\odot$ \citep[e.g.][]{cattaneo06, gabor10}.

\item We restrict the heating to within a certain radius
  $r_{\rmn{heat}}$ from the center of the halo.  This is an attempt to
  mimic feedback from a central radio AGN.  \citet{voit05} argue,
  based on observations of an entropy floor within the central 30 kpc
  in cooling flow clusters from \citet{donahue06}, that an AGN
  outburst sufficient to counteract the cooling will create subsonic,
  buoyant bubbles at a typical radius of 30 kpc.  The region outside
  this radius should be heated by the expanding bubbles.  Following
  this, we use $r_{\rmn{heat}} = 30$ kpc.  A numerical difficulty with
  this is that we found that large galaxies in our simulations have
  star-forming regions at significantly larger radii.  This result
  occurs for galaxies identified using both {\sc FOF} and {\sc SKID}
  methods.  This may owe either to satellite galaxies at these radii
  being grouped into the central galaxy, or to the threshold density
  for star formation being too low.  We note that the SF threshold is
  $\approx 0.1$~cm$^{-3}$, which is well below ISM densities where
  star formation occurs in reality.  This is a historical numerical
  convenience motivated by the inability of our simulations to fully
  resolve ISM densities.  In any case, this variant
  produces a red sequence, but has too many massive galaxies remain
  blue due to the ongoing star-formation.  Apparently in our
  simulations, heating only the central regions of a halo does not
  heat the outer regions as suggested by observations of sound waves
  in clusters \citep{fabian06}.  The failure of central heating to
  spread energy to large radii may suggest that we do not adequately
  resolve the hydrodynamics in our typical hot haloes, or that raising
  the temperature to only $T_{\rm vir}$ may be insufficient, and we
  need to raise it further.

\item We do not heat gas below our 250000 K critical temperature.
This is intended to allow for cold flows from IGM filaments to
penetrate into hot haloes, as has been argued by~\citet{dekel09}.  We
discuss this variant further in \S\ref{sec.variations}.

\item We do not heat gas within sub-haloes.  As indicated by Figure
  \ref{fig.fof_halomass}, many small satellite galaxies live in the
  hot gas haloes of their larger hosts.  AGN in these small galaxies
  may or may not contribute significantly to the overall heating of
  the hot gas halo.  Since their importance in quenching massive
  galaxies is debatable, we implemented a method to identify these
  sub-haloes and exclude them from the feedback prescription.  When
  calculating hot gas fractions as described above, we estimate the
  virial radius of each {\sc FOF} galaxy.  If a galaxy's center falls
  within the virial radius of another, more massive, galaxy, then we
  consider it a sub-halo.  Once we identify sub-haloes, we do not
  apply the thermal feedback to gas within them.  We discuss results
  of this mechanism in \S\ref{sec.variations}.

\end{enumerate}

\subsection{Simulation parameters}

Simulating massive galaxies presents a substantially larger
computational challenge than simulating more common systems.  One is
pushed towards larger volumes ($\ga 50$~Mpc/h) by the need to obtain a
sufficient statistical sample of massive galaxies.  Conversely, since
massive galaxies begin forming stars first, this requires simulations
that robustly resolve early star formation, requiring higher
resolution ($\sim$few~kpc).  Given computational resource limitations,
we can only perform a few large runs.  To explore all the variants
listed above, we are forced to run smaller simulations.  We choose to
compromise on volume to ensure robust star formation histories.

Our primary simulations use a box size of $48 h^{-1}$ Mpc and a
gravitational softening length of $3.75 h^{-1}$ kpc (equivalent
Plummer), with 256$^3$ dark matter and $256^3$ gas particles.  This
yields a gas particle mass of $1.2\times 10^8 M_\odot$, and a dark
matter particle mass $5.1\times$ larger.  Our smaller simulations keep
the same mass and spatial resolution, but reduce the box size to $24
h^{-1}$ Mpc with 128$^3$ dark matter and 128$^3$ gas particles.  All
simulations and analyses use a \emph{Wilkinson Microwave Anisotropy
Probe} concordance cosmology \citep{komatsu09} with $H_0 \equiv 100h =
70$km s$^{-1}$ Mpc$^{-1}$, matter density $\Omega_m=0.28$, baryon
density $\Omega_b=0.046$, a cosmological constant with
$\Omega_{\Lambda} = 0.72$, root mean square mass fluctuation at
separations of 8~Mpc $\sigma_8 = 0.82$, and a spectral index of
$n=0.96$.  Our small runs are useful for testing some of the
variations in the quenching physics, and also for tracking details of
each on-the-fly galaxy and feedback event.  The larger simulations do
a much better job of sampling the massive end of the luminosity
function, but take many weeks to run.

We focus primarily on three models which we refer to as {\it no
quenching, merger quenching, and hot gas quenching.}  The no quenching
model includes star formation and momentum-conserving winds, but no
additional explicit feedback mechanism.  It is effectively the same
model as the ``vzw'' model of \citet{oppenheimer08}.  The merger
quenching model incorporates 1500 km s$^{-1}$ winds from merger
remnants to quench star formation.  For the hot gas quenching, we
focus on a model where we heat only the non-star-forming gas particles in
{\sc FOF} galaxies with $f_{\rmn{hot}}>0.6$.  In \S \ref{sec.results}
we will also explore some simulations with slight variations of the
quenching models.

\subsection{Simulation outputs and analysis tools}

We output snapshots of each simulation at 108 redshifts, starting at
$z=30$ and ending at $z=0$.  The snapshots contain information for
every simulation particle, including position, velocity, mass,
metallicity, gas density, gas temperature, star-formation rate, and
time of formation (for star particles).  In addition to these outputs,
we save the galaxy catalog from each instance of the on-the-fly galaxy
finder for later analysis.

Our suite of analysis tools allows us to compare our simulation
results directly with observations.  {\sc SKID} provides a list of member
particles (star and star-forming gas) for each simulated galaxy.  The
sum of star particle masses is then the galaxy stellar mass, and the
star formation rates of the gas particles are summed to give the star
formation rate of the galaxy.

We then calculate galaxy spectra using the models of \citet{bc03}.  We
treat each star particle as a single stellar population with an age
and metallicity determined directly in the simulation.  By adding up
the spectra of all star particles within a galaxy, we obtain the
spectrum of that galaxy.  Then we convolve galaxy spectra with
observational filter curves to obtain broad-band colors and
magnitudes.

\section{Observational Constraints} \label{sec.observations} 

To characterize the effectiveness of our quenching models, we compare
the results of our simulated galaxy populations to galaxy observations
in the low-redshift ($z<0.1$) universe.  Our goal is to create a
realistic red sequence, so we focus on color-magnitude diagrams (CMDs)
and red-galaxy luminosity functions (LFs).

We follow \citet{gabor10} in using the low-redshift version of the
Value-Added Galaxy Catalog \citep[VAGC, ][]{blanton05_vagc} of the
SDSS \citep{adelman-mccarthy08, padmanabhan08}.  This catalog includes
$ugriz+JHK$ absolute magnitudes for $\sim 170000$ galaxies with
$z\lesssim0.05$.  After converting absolute magnitudes to our
preferred cosmology (with $h=0.70$), we can straightforwardly plot
CMDs.  We will plot $g-r$ vs $r$ because the $r$-band light is a good
tracer of stellar mass.  For the luminosity functions, we use $r$-band
luminosities and the $1/V_{\rmn{max}}$ method \citep{schmidt68} with
the $V_{\rmn{max}}$ values in the VAGC.  We compare these
observational results with simulation snapshots at $z=0.025$, which
we will refer to as low redshift or $z=0$.

As discussed in \citet{gabor10} and \S\ref{sec.blue_galaxies},
uncertainties in dust obscuration motivate us to compare mass
functions (MFs) of blue galaxies rather than luminosity functions.
For this we use publicly available stellar masses derived for SDSS
galaxies by fitting observed galaxy spectra to templates
\citep{kauffmann03_stellarmass_measurements}.  We cross correlate this
sample with the VAGC to obtain $\sim 40000$ galaxies with stellar
masses and broad-band colors for the MFs.

\section{Constraining Quenching Models}
\label{sec.results}

\subsection{Basic models: Red Galaxies}

\begin{figure}
\includegraphics[width=84mm]{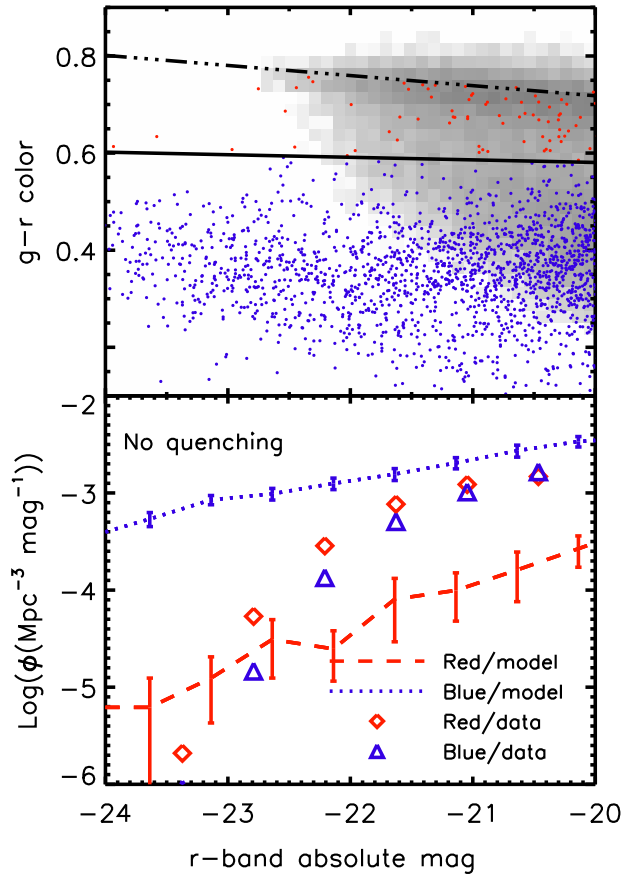}
\caption{CMD (top panel) and LF (bottom) for a simulation with no
  quenching.  In the CMD, grayscale represents the density of SDSS
  galaxies, points are simulated galaxies, the dot-dashed line is a
  fit to the observed red sequence, and the solid line separates the
  simulated red sequence and blue cloud.  Without quenching, our
  simulations do not produce red galaxies.  In the LF, symbols
  represent the observed luminosity functions for red and blue
  galaxies, and lines are for simulated galaxies.  The red galaxy
  luminosity function is too low by roughly an order of magnitude.
  Simulated blue galaxy luminosities are computed without a correction
  for dust to distinguish the intrinsic red sequence more easily.}
\label{fig.cmdlf_noquench}
\end{figure}

\begin{figure}
\includegraphics[width=84mm]{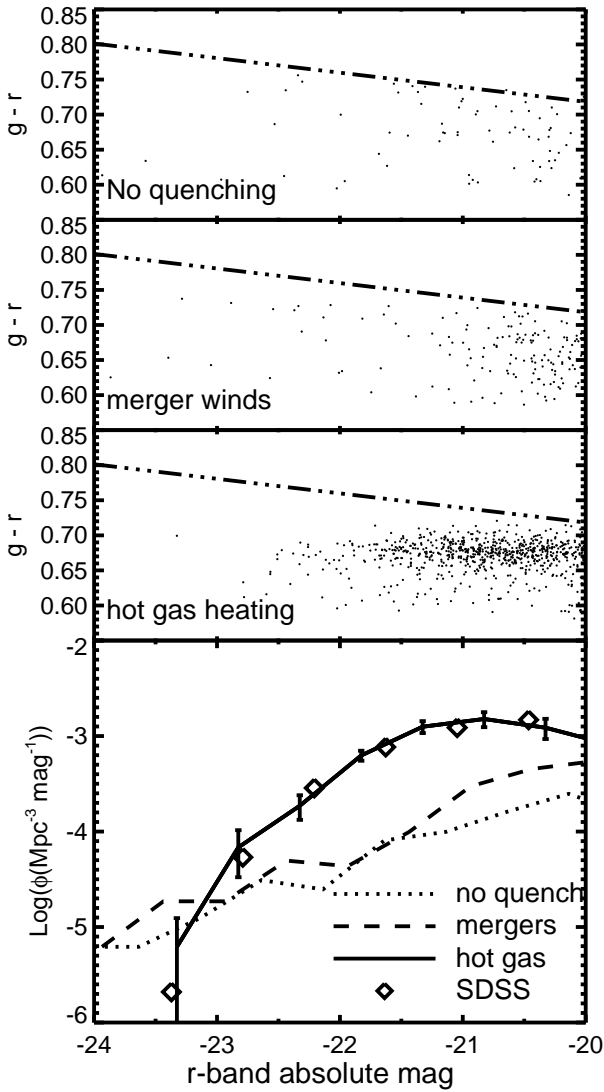}
\caption{CMDs zoomed in on the red sequence (top 3 panels) and LFs
  (bottom panel) for simulations with no quenching, merger quenching,
  and hot gas quenching.  This shows the main result of this work:
  merger quenching fails to produce a substantial red sequence,
  whereas hot gas quenching succeeds.}
\label{fig.cmdlf_main}
\end{figure}

\begin{figure}
\includegraphics[width=84mm]{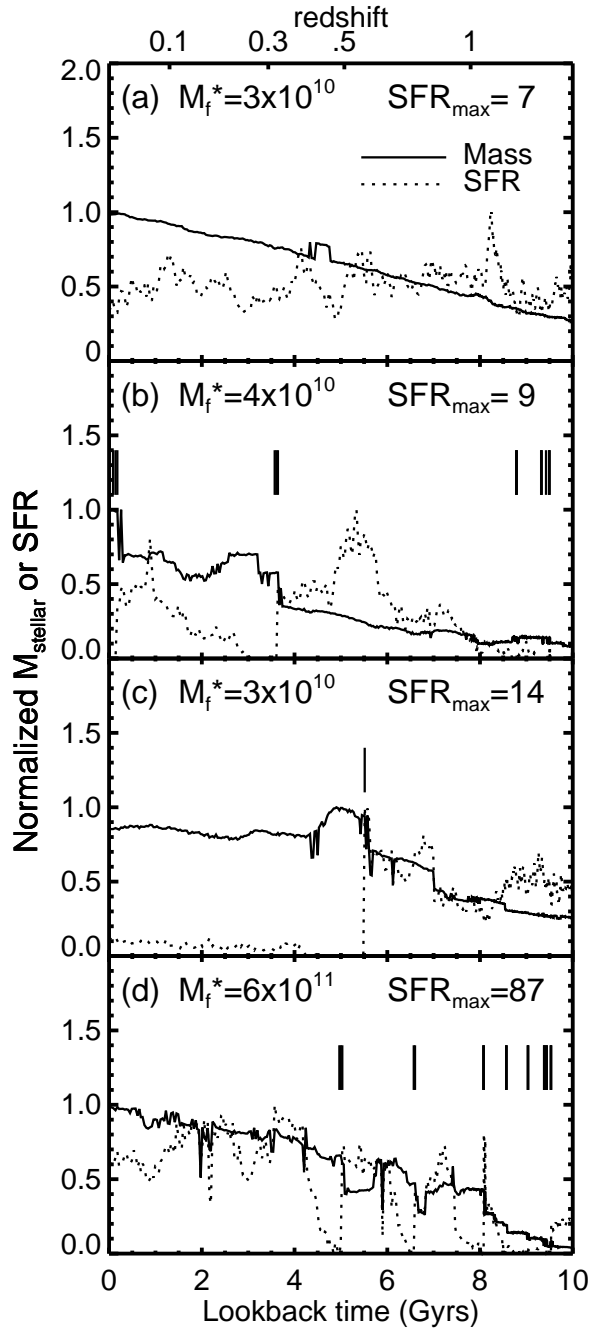}
\caption{Cosmic histories of mass (solid lines) and instantaneous
  star-formation rate (dotted lines) for 4 galaxies selected from a
  simulation with merger quenching.  These illustrate the variety of
  outcomes following major mergers.  Vertical solid lines above mass
  histories indicate major mergers, and we show the $z=0$ stellar mass
  ($M_f^*$ in $M_{\sun}$) and maximum star-formation rate (in
  $M_{\sun}$yr$^{-1}$) for each.  (a) A galaxy with moderate, fairly
  constant star-formation rate.  (b) A galaxy with some major mergers.
  By design, the star-formation rate drops to zero immediately
  following a merger.  The star-formation rate recovers within about 2
  Gyrs due to gas accretion and possible minor mergers.  (c) A galaxy
  whose major merger quenches almost all subsequent star formation
  until $z=0$.  (d) The complex mass history of the central galaxy in
  one of the two most massive groups in the simulation.  This galaxy
  undergoes several major mergers, after each of which the SFR
  recovers within 1 Gyr due to fresh inflows of gaseous fuel.}
\label{fig.sfr_hist}
\end{figure}

\begin{figure}
\includegraphics[width=84mm]{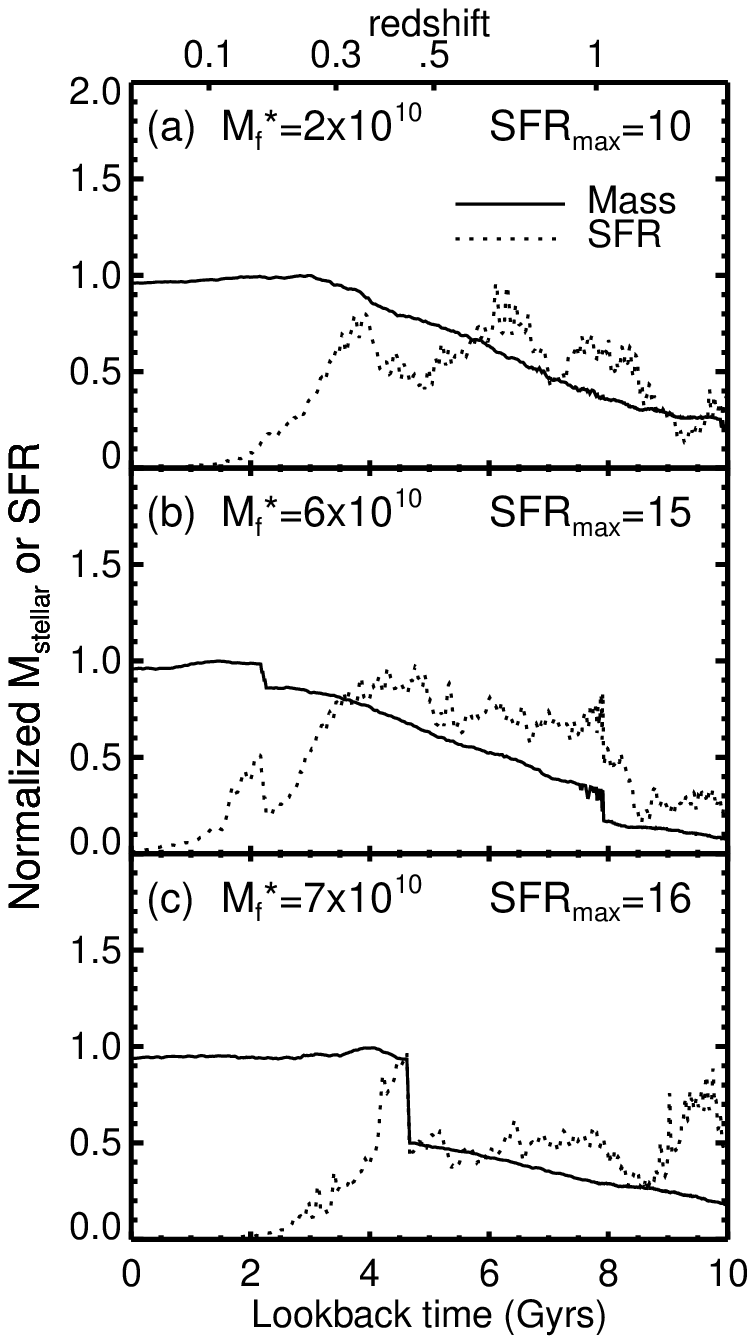}
\caption{Star formation histories (dotted lines) and stellar masses
  (solid lines) for three galaxies in our hot gas quenching model
  (cf. Figure \ref{fig.sfr_hist}).  (a) A galaxy with fairly quiescent
  star-formation and steady mass growth.  At $z\approx 0.3$, this
  galaxy's halo becomes dominated by hot gas, after which we begin
  heating the gas to prevent cooling.  Starved of new fuel,
  star-formation declines to zero over $\sim 2$ Gyr.  (b) A minor
  merger (lookback time $\approx 2$ Gyr) delivers fuel to a galaxy
  which had begun quenching about 2 Gyr earlier.  (c) A major merger
  increases a galaxy's halo mass to the point where it can support a
  hot halo, after which it is quenched.}
\label{fig.sfr_hist_hfof}
\end{figure}

Since our goal is to reproduce a realistic red sequence, we compare
the red galaxy CMDs and LFs produced by our various models to
determine the best match.  We first illustrate the problem with
simulations that do not have additional quenching mechanisms.  Figure
\ref{fig.cmdlf_noquench} shows the full CMD and LF for a simulation
with no quenching feedback, compared to observations.  Following
\citet{gabor10}, we separate red simulated galaxies from blue in the
CMD using the solid line.  This line is bluer and shallower than that
used for observed galaxies because it better captures the bimodality
in our most successful models.  As we discuss below \citep[and in
][]{gabor10}, our model red sequence is generally too blue and has
little or no slope.  We estimate LF uncertainties based on a jackknife
re-sampling of the simulation volume \citep{finlator06} (typical
observational uncertainties are smaller than the symbols except for
the brightest bins).

Almost all galaxies in this no-quenching simulation constantly accrete
gas from the IGM, and this gas supplies the fuel for ongoing star
formation.  As a result, the simulation produces few red galaxies and
many very massive, very bright blue galaxies.  The LF reflects this
effect.  The red galaxy LF is too low (relative to that observed) by
an order of magnitude for $r\approx -21.5$, and the blue LF is too
high.  Furthermore, this model does not produce the observed sharp
turnover in the luminosity function at high
masses~\citep{oppenheimer10}.

We show the simulated blue galaxies without accounting for obscuration
from dust to illustrate the intrinsic galaxy luminosities.  Dust
obscuration will push the blue galaxies up and to the right (redder
and fainter) in the CMD, possibly contaminating the red sequence.  The
observed universe does not contain star-forming galaxies as massive as
those shown here, so estimating the extinction in these simulated
galaxies would require large extrapolations that are unlikely to be
robust.  We discuss dust extinction and blue galaxies further in
\S\ref{sec.blue_galaxies}.

Focusing just on the red sequence, we compare results for models with
no quenching, merger quenching, and hot gas quenching in Figure
\ref{fig.cmdlf_main}.  Like the no quenching simulation, the merger
quenching model leaves a very sparse red sequence.  Our hot gas
quenching model, however, produces a red sequence of galaxies
well-separated from the blue cloud.

The luminosity function panel in Figure \ref{fig.cmdlf_main}
represents one of the central results of this paper.  It emphasizes
that our hot gas quenching model nicely produces a sharp truncation in
the red galaxy LF, while our merger quenching model does not.  Because
it produces so few red galaxies, merger quenching does not
significantly improve the red galaxy luminosity function of our
simulations without quenching.  The hot gas quenching model fares
well, matching the observed red sequence luminosity function at all
magnitudes we resolve.

For the very brightest galaxies, the relatively small volume of our
simulations hampers the comparison with data -- we do not effectively
sample the highest peaks in the density distribution that lead to the
most massive galaxies.  This is reflected in our errorbars.

Examining the halo quenching red sequence further, we see that it is
too blue by about 0.1 magnitudes in $g-r$, and has a nearly flat slope
that fails to match the observed trend that brighter red galaxies are
redder.  We examined this issue at length in \citet{gabor10}, and the
same underlying causes are likely here.  The overall ``blueness
problem'' may result from either metallicities that are too low or
ages that are too young.  Our red galaxies have metallicities lower
than those inferred for the observed red sequence in the VAGC, and
lower metallicities lead to bluer stars.  Mean stellar ages of our
galaxies are at least as old as those inferred from the observations,
so incorrect ages cannot cause the red sequence to be too blue.  Since
we match the stellar masses, the metallicity deficit is unlikely to
owe to an underproduction of stars.  It could be that our outflow
model ejects too many metals from these systems, or else we might have
adopted supernova metal yields that are too low.  The red sequence slope poses
a subtler problem, perhaps suggesting that simulated massive galaxies
merge with too many, too massive, or too young satellite galaxies with
lower metallicities.  In \S\ref{sec.blue_galaxies}, we describe a
metallicity re-calibration that matches our simulated red galaxy
colors to those observed.

Why does merger quenching fail?  In \citet{gabor10} we showed that if
star formation is completely quenched in the remnants of major
mergers, then our models can produce a reasonable match to the red
sequence.  Merger quenching therefore does not fail because there are
too few major mergers.  Rather, merger quenching fails because star
formation resumes even after a merger remnant loses all its
star-forming gas.

We illustrate the resumption of star formation after merging in Figure
\ref{fig.sfr_hist}.  Each panel corresponds to an individual galaxy
taken from a small test simulation with merger quenching.  For each
galaxy we show the normalized star-formation history (dotted lines)
and stellar mass history (solid lines) over cosmic time, with vertical
tick marks indicating major mergers.  Galaxies with no major mergers
(panel a) have fairly constant star formation histories (since
$z\sim2$), with peaks and valleys depending on the details of
accretion from the IGM and outflows powered by stars.  Galaxies that
undergo major mergers (panels b--d) generically stop forming stars
when a merger occurs because we instantaneously eject all the cold
gas.  However, in every case the merger remnant resumes star formation
as new fuel is accreted from the IGM.  The timing and intensity of the
resumption of star-formation varies, probably depending on
environment.  A very massive galaxy living at the center of a large
potential well (panel d) will accrete new gas and return to high
star-formation rates in $\sim 1$ Gyr.  Some smaller galaxies
(e.g. panel c) never return to SFRs near their pre-merger levels,
though they do resume some star-formation within $\sim 2$ Gyrs.

As a comparison, we show star-formation histories for quenched
galaxies in our hot gas quenching model in Figure
\ref{fig.sfr_hist_hfof}.  The galaxy in panel (a) grows steadily in
stellar mass until its halo becomes dominated by hot gas (at $z\approx
0.3$).  At this point our hot gas heating cuts off additional fuel for
star-formation, and the galaxy gradually uses up its pre-existing
reservoir of gas.  Star formation peters out after $\sim 2$ Gyrs.
Even after the starvation process begins, merging galaxies can
momentarily supply new fuel, as reflected by a jump in SFR in panel
(b) at a lookback time of $\sim 2$ Gyrs.  In panel (c) we show a
galaxy whose major merger raises its halo mass enough that a hot gas
halo becomes stable.  These star-formation histories all show a slow
decline in star-formation rate over $\sim 2$ Gyr timescales, similar to the star-formation timescale in our sub-grid star-formation
model \citep{springel03}.  After
star formation stops, the galaxy stellar mass declines slowly owing to
stellar evolution.

In summary, a simple model where all star formation is quenched in
galaxies with a hot halo fraction above 60\% produces a red galaxy
luminosity function in very good agreement with observations.  It also
produces a reasonable red sequence, although it is still too blue and
too shallow as we saw previously in \citet{gabor10}, likely reflecting
issues with enrichment.  In contrast, assuming that all major mergers
are associated with 1500 km s$^{-1}$ outflows as suggested by
observations \citep{tremonti07} does not significantly populate the
red sequence.

\subsection{Basic models: blue galaxies}
\label{sec.blue_galaxies}
\begin{figure}
\includegraphics[width=84mm]{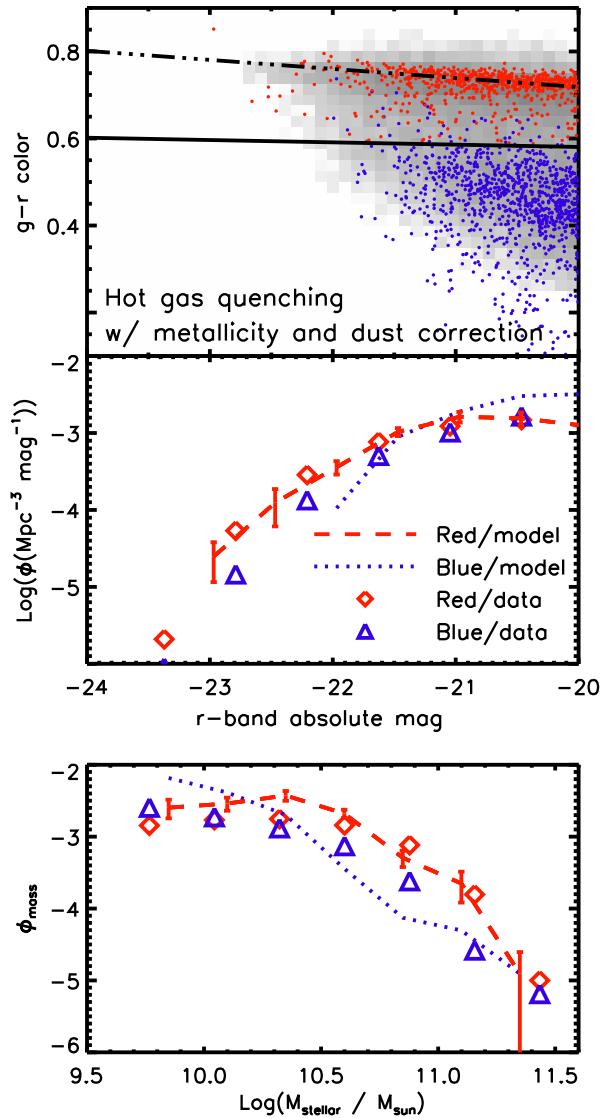}
\caption{Full CMD (top panel), LFs (middle panel), and stellar mass
  functions (lower panel) for our hot gas quenching model.  The
  simulated CMD includes a metallicity re-scaling to best reproduce
  the observed red sequence colors, as well as dust reddening and
  extinction for the blue galaxies.  Due to uncertainties in the dust
  prescription, we prefer to use stellar mass functions (lower panel)
  to compare simulated and observed blue galaxies.  Our hot gas
  quenching model yields too few massive blue galaxies.  }
\label{fig.cmdlf_mf_hfof}
\end{figure}

\begin{figure}
\includegraphics[width=84mm]{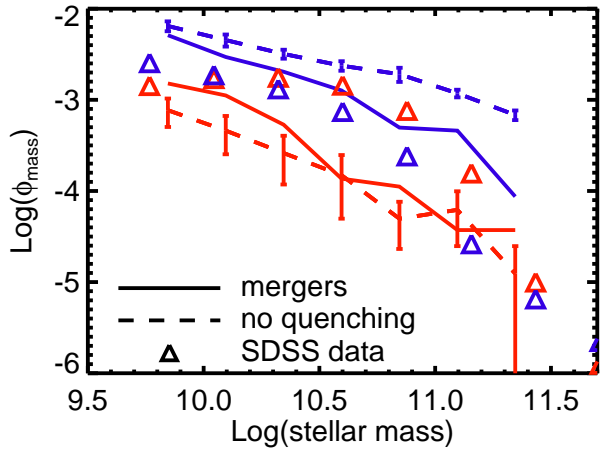}
\caption{Stellar mass functions split into blue and red galaxies
  (colors) for a simulation with no quenching (dashed lines), a
  simulation with merger quenching (solid lines), and SDSS data
  (triangles).  Inducing powerful outflows after galaxy mergers does
  suppress the growth of massive blue galaxies, but it does little to
  increase the number of red sequence galaxies above that produced in
  the no-quenching simulation.}
\label{fig.mf_m15_noquench}
\end{figure}


We have shown that additional heating in hot gas halos can produce a
red sequence whose luminosity function matches that observed in the
local universe.  We now examine galaxy colours more closely.  Since
these are sensitive to uncertain dust corrections and problems with
metallicities, we will use galaxy stellar mass functions later in this
section.


Figure \ref{fig.cmdlf_mf_hfof} shows a complete CMD and corresponding
LFs for our hot gas quenching model.  Here, we apply an ad hoc
metallicity correction to force the simulated red sequence to match
the observed colors.  This assumes that our predicted stellar ages are
correct, but that our simulations underpredict stellar metallicities
in massive red galaxies.  We add a metallicity of $Z_{\rmn{add}}$ to
each single stellar population (i.e. star particle) when calculating
galaxy colors using stellar population synthesis models.  By trial and
error, we find a good fit using the relation
\begin{equation}\label{eq:zadd}
Z_{\rmn{add}}=0.003 + (6\times10^{-14} M_{\rmn{stellar}} / M_{\sun}).
\end{equation}
These are absolute metal mass fractions, and solar metallicity here is
$\approx0.012$ \citep{asplund05}.  This correction gives an indication
of how far off our metallicities must be.  For a $10^{11}M_{\sun}$
galaxy (with typical absolute magnitude $r\sim -22$), $Z_{add} =
0.009$, representing an increase of $\sim50$\% over typical simulated
metallicities for such galaxies.

We show blue galaxies in the CMD with a correction for dust extinction
of optical light in the $g$ and $r$ bands.  In our dust prescription,
the $E(B-V)$ extinction scales with star-formation rate based on
empirical relations with UV luminosity of local galaxies
\citep{wang96, somerville01, finlator06}.  We assume a
\citet{calzetti00} reddening law.  This dust prescription yields a
blue cloud in fairly good agreement with data, especially in the
qualitative shape.  It also shifts a small number of intrinsically
blue galaxies onto the red sequence, but we count these galaxies as
blue for the purposes of calculating the luminosity and mass
functions.

In \citet{gabor10}, we discussed the difficulties in modeling
extinction.  Given a model galaxy with a specified star-formation and
gas accretion history, we need to know the level of dust extinction
and reddening to compare colors with observed galaxies.  Several dust
extinction prescriptions keyed to physical properties of galaxies,
such as metallicity or star-formation rate, have been proposed
\citep{finlator06}.  Different dust prescriptions yield substantially
different colors and magnitudes for our star-forming model galaxies,
suggesting large uncertainties.  This is especially true for large,
bright blue galaxies since these are not well represented in the real
Universe.  For red galaxies this is not a problem because they have
very little extinction \citep{lauer05}.

One way to mitigate the large uncertainties in dust obscuration is by
examining stellar mass functions for the blue galaxies, rather than
luminosity functions.  We use stellar mass estimates from
\citet{kauffmann03_stellarmass_measurements} based on SDSS spectra,
which account for extinction self-consistently within the SED fitting
procedure for each galaxy.  We cross-correlate these stellar mass
measurements with galaxies from the VAGC, allowing us to use
broad-band colors and generate mass functions using the
$1/V_{\rmn{max}}$ method.  Using the broad-band colors, we separate
the galaxies into the red sequence and blue cloud as before.

We show the resulting stellar mass functions for our hot gas quenching
model in the bottom panel of Figure \ref{fig.cmdlf_mf_hfof}.  As
expected from the luminosity function results, and given the tight
correspondence between luminosity and mass for old stellar
populations, the red galaxy mass function matches the observations
well.  The blue galaxy stellar mass function is overall a good match
to data as well, although it shows a slight underproduction of
star-forming galaxies just above the knee of the LF.  The small number
of galaxies ($<10$) populating the bright end show only trace levels
of star-formation.  Our hot fraction criterion ($f_{\rmn{hot}}>60$\%)
for feedback results in a fairly sharp critical halo mass of
$\sim10^{12}M_{\sun}$ (Figure \ref{fig.fof_hotfrac}), and therefore
stellar mass of $\sim 10^{10.5}M_{\sun}$, above which fueling of
star-formation is turned off.  Nevertheless, later infall of blue
systems can turn a massive galaxy somewhat bluer temporarily.  Hence
this model produces massive blue galaxies, although the details of
that are quite sensitive to our quenching prescription and should be
taken with some caution.

This quenching model minimally impacts small galaxies, basically only
when they are satellites within larger halos.  \citet{dave11} show
that the fraction of satellite galaxies at these masses is about
one-third, which explains why any impact on satellites results in only
minor changes to the mass function.  At the faint end, our model
somewhat overproduces the faint star-forming galaxy population, which is a
result also seen in the total mass function examined in \citet{oppenheimer10}.
Thus, while our hot gas quenching model successfully matches the
observed number densities of red sequence galaxies, it fails to match
all the details of the blue galaxy population.

For comparison, we show the galaxy stellar mass functions for our
no-quenching and merger outflows simulations in Figure
\ref{fig.mf_m15_noquench}.  Solid lines correspond to our no-quenching
model, and dashed lines to merger quenching.  By ejecting the gas from
galaxy merger remnants, we substantially reduce the number density of
very massive star-forming galaxies.  As indicated by the luminosity
function analysis above, merger quenching does not substantially
increase the number of red galaxies.  In both models, both red and
blue mass functions look roughly like power laws, lacking the
exponential cutoff observed in SDSS.  This shows that merger quenching
helps with only one of the two problems in the simulated LFs, namely
that it reduces the number of blue galaxies, but it does not increase
the number of red galaxies.

In summary, quenching by halo hot fraction successully reproduces the
SDSS red galaxy LF, and broadly reproduces the SDSS red sequence.  An
empirical augmentation to galaxy metallicities given in
equation~\ref{eq:zadd}, which increases metal content by typically
50\% (with a range of 10--100\%), results in an excellent match to the
red sequence.  The predicted red galaxy mass function is likewise an
excellent fit to data, and the blue galaxy mass function broadly
matches as well although there is a hint of too few bright galaxies
and too many faint ones.  The agreement using hot halo quenching is
far superior to that using merger quenching, as the latter does not
increase the red sequence over a no-quenching model although it does
significantly suppress blue galaxies.  These simulations therefore
suggest that halo gas quenching, by itself, drives the formation of
the red sequence.

\subsection{Model variations}
\label{sec.variations}
\begin{figure}
\includegraphics[width=84mm]{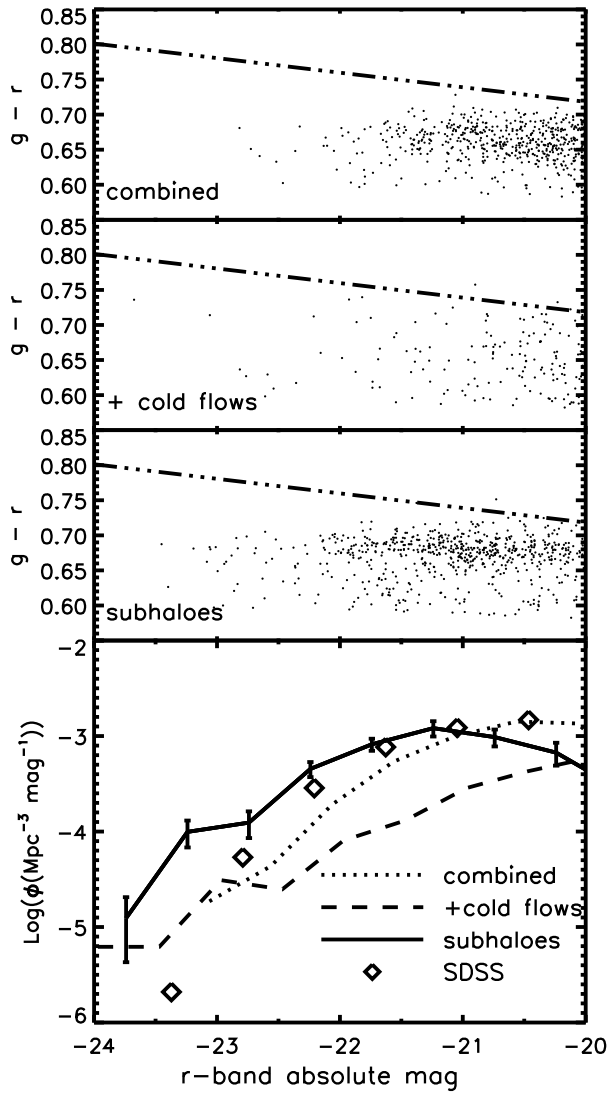}
\caption{CMDs (top 3 panels) and LFs (bottom panel) for three
  alternative quenching models: combined hot gas heating $+$ merger
  winds (top CMD, dotted line in LF), the same combined model except
  \emph{without} heating any gas below 250000 K (middle CMD, dashed
  line in LF), and a model with hot gas quenching \emph{without}
  heating gas around sub-haloes (bottom CMD, solid line in LF).  The
  combined model performs well, with slightly lower LFs than the hot
  gas quenching-only model.  Allowing cold flows fails due to wind
  recycling, and allowing gas fueling in sub-haloes leads to
  discrepancies at both low and high luminosities in this range.}
\label{fig.cmdlf_combo}
\end{figure}

\begin{figure}
\includegraphics[width=84mm]{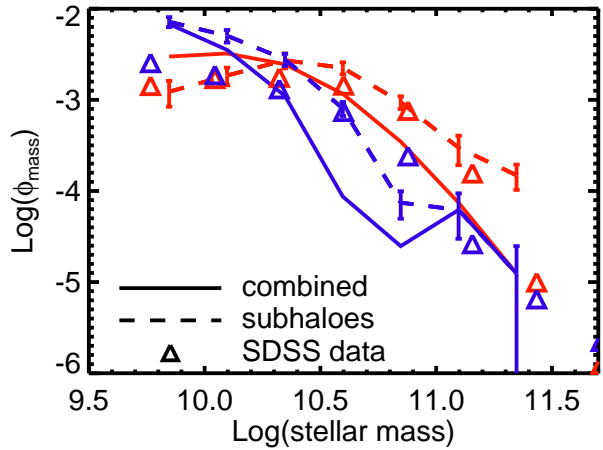}
\caption{Blue and red galaxy mass functions (color-coded) for our
  combined quenching model (solid lines) and our model without heating
  gas around sub-haloes (dashed lines).  The combined model shows a
  larger deficit of massive blue galaxies, and the subhaloes model
  shows a similar deficit to that in our basic hot gas heating model.}
\label{fig.mf_combo}
\end{figure}

We now consider physically-motivated variations on our basic models.
Here we present three variations that all include hot gas heating.  In
the first we combine our merger winds quenching model with our hot gas
heating model.  Both of these mechanisms are well-motivated by
observations, hence both processes may occur in the real universe.  We
refer to this as the ``combined'' model.

In a second variation, we add an exception to the hot gas heating
model.  Several authors have argued that cold flows of gas along
filaments can penetrate hot gas haloes at high redshift, since
filaments are denser at earlier epochs.  In principle our simulations
should capture this behaviour with their hydrodynamic treatment.  Our
basic hot gas heating model, however, bypasses the hydrodynamics to
heat all (non-star-forming) gas around a galaxy.  In order to prevent
artificial heating of these cold flows, allowing them to penetrate hot
haloes, we implemented a model where we only heat gas that is already
above our hot gas cutoff of 250000 K.  We present this variation with
the combined mergers $+$ hot gas model, but we note that the results
are similar when we apply this variation to our basic hot gas heating
model.  We refer to this as the ``cold flows'' model.

In a third variation, we tested a model where we do not heat the hot
gas in the vicinity of sub-haloes.  In our simulations, some
sub-haloes are embedded in their parent's hot gas halo, even if the
sub-halo is too small to support its own hot gas.  Our basic quenching
model heats the gas in these sub-haloes just the same as in the
massive, parent haloes.  Heating from AGN feedback is typically
observed in the centers of groups and clusters, and it is not clear
whether heating from AGN in small satellite galaxies actually occurs
in the real Universe.  In our simulations, many satellites
particularly at low masses are stripped of gas owing to ram pressure
or starvation \citep{dave11}, but larger satellites are typically
star-forming.  X-ray observations do suggest that most satellites in
clusters have mini cooling cores analogous to those seen in brightest
cluster galaxies, and that heating by supernovae or AGN may be
necessary to quench them \citep{sun07}.  Some satellite elliptical
galaxies show evidence for AGN outbursts \citep{jones02, machacek06}.

To distinguish sub-haloes (or satellite galaxies) from their parent haloes,
we use the virial radii estimated when calculating hot gas fractions.
If we consider two haloes A and B, then halo A is a subhalo of halo B
if and only if (1) the center of halo A falls within halo B's virial
radius and (2) halo B is more massive than halo A.  We flag sub-haloes
in our group catalog, and we do not add heat to their gas.  Note that
in this "subhaloes" variant, we do not induce outflows in galaxy
merger remnants.

We show results from these three model variations in Figures
\ref{fig.cmdlf_combo} and \ref{fig.mf_combo}.  Figure
\ref{fig.cmdlf_combo}, analogous to Figure \ref{fig.cmdlf_main}, shows
CMDs and LFs for just the red sequence in our three variant quenching
models.  Combined quenching has only a small overall effect on the red
sequence when compared to our basic hot gas quenching: at all masses,
the number density of red galaxies is slightly below that observed.
In short, merger quenching has little overall effect on the galaxy
population.

Our ``cold flows'' model fails to produce a red sequence.  This is
\emph{not} due primarily to pristine infall along filaments, but
rather due to recycling of material expelled by star formation-driven
winds \citep{oppenheimer10}.  Wind recycling dominates infall in
galaxies since $z\sim 1$, and is stronger in more massive systems that
have short recycling times.  Although particles in these systems are
ejected at speeds of up to $\sim 1000$~km/s, their cooling
times times are short owing to the high level of enrichment, and so
many never heat above the shock threshold.  We discuss this issue
further in \S \ref{sec.discussion}.

Our variation that excludes heating in sub-haloes (the ``subhaloes''
model) increases the number of bright red galaxies and decreases the
number of faint ($r\la -20$) red ones.  The deficit of faint red
galaxies is straightforward -- turning off quenching in satellites
allows those satellites to accrete gas and remain blue.  The excess of
bright red galaxies arises because the massive galaxies that make up
the bright end grow significant mass via mergers with their
satellites, even after the central galaxies have been quenched
\citep{gabor10}.  Without quenching, these merging satellites form
more stars before they are subsumed than they do with quenching.
These mergers then result in more massive central galaxies, and a
corresponding increase in the bright-end luminosity function.

To study the blue galaxies in these variant models, we again look at
stellar mass functions.  Figure \ref{fig.mf_combo} shows mass
functions for the combined quenching model (solid) and the hot gas
quenching model without sub-halo heating.  In comparison to our basic
hot gas heating model, the combined model results in a sharper cut-off
in the blue galaxy mass function above $M_{\rmn{stellar}}\sim
10^{10.5} M_{\sun}$, leading to a larger deficit of massive blue
galaxies.  The model without sub-halo quenching yields a similar blue
galaxy mass function to our basic hot gas heating model.

Overall, the variant models we tested do not solve the underlying
problems with our basic models, while introducing new problems.  Our
combined model exacerbates the deficit of blue galaxies at masses
$\gtrsim 10^{10.5} M_{\sun}$.  Our model with cold flows fails to
produce a red sequence.  Our model without heating of sub-haloes
creates discrepancies in the bright and faint end red galaxy LF.

The combined model leads to similar results as the basic model, but
with fewer red sequence galaxies.  In \citet{gabor10} we showed that
the halo mass cutoff, or correspondingly the hot fraction cutoff,
above which we quench star-formation largely determines where the
exponential cutoff begins in the LF.  We therefore expect that raising
$f_{\rmn{hot}}^{\rmn{crit}}$ from 0.6 to, say, 0.7 might bring the
combined quenching model into better agreement with observations.  We
cannot therefore easily distinguish between the basic hot gas
quenching and the combined quenching models.

In summary, by testing many variants on our basic quenching models, we
find that hot gas quenching is required to get a red sequence, while
merger quenching has minimal impact.  Variations based on not
quenching gas in cold flows or subhalos have a noticeable but not
large effect on the LF and CMD.  None of these variants impact the
slope or blueness of the red sequence, which continue to be a
difficulty.  Hence while all these quenching model variants are
physically motivated, in the end a model with simple hot gas quenching
matches the ensemble data as well as or better than any of the others.

\subsection{The dependence of merger quenching on the critical mass ratio}
\label{sec.appendix}
\begin{figure*}
\includegraphics[width=160mm]{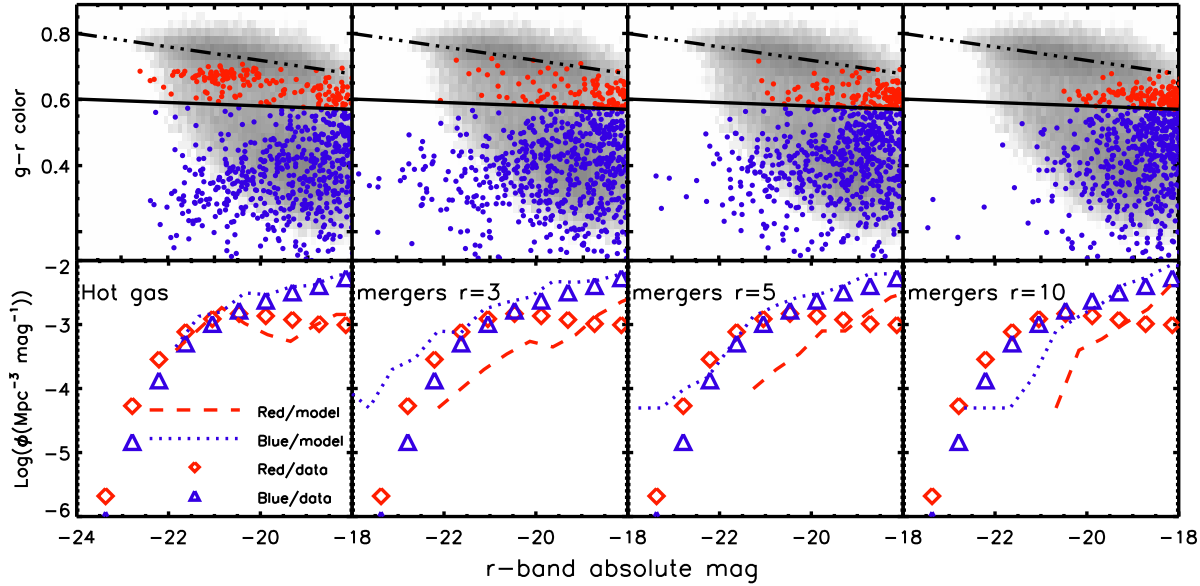}
\caption{Using a critical mass ratio $r>3$ to define major mergers
does not yield a red sequence, as indicated by colour-magnitude
diagrams (top row) and luminosity functions (bottom row) for hot gas
quenching (left column) and merger quenching with $r=3,5,$ and $10$.
These simulations were run in boxes of side length $24 h^{-1}$ Mpc.
Even at this smaller volume, hot gas quenching results in an obvious
red sequence with number densities close to those observed.  Merger
quenching with values of $r>3$ leads to a stronger suppression of
galaxy growth (fewer bright galaxies, e.g. right column), but does not
yield a distinct red sequence with substantial number densities.}
\label{fig.appendix}
\end{figure*}

Studies typically define major mergers as those with a 3:1 mass ratio
or smaller.  We have adopted this critical mass ratio ($r=3$) here for merger
quenching, assuming that mergers with larger mass ratios have no
effect.  In this appendix, we show that allowing larger mass ratio
mergers to trigger quenching does not lead to a red sequence in better
agreement with observations.  Although using $r>3$ leads, in a general sense, to more quenching, it does not lead to a bimodality in galaxy colors nor a significant number of red galaxies.

Figure \ref{fig.appendix} compares color-magnitude diagrams and
luminosity functions for four different simulations: one with our
preferred hot gas quenching model, and our merger quenching model with
three values of the critical mass ratio, $r=3, 5,$ and $10$.  These
simulations are run with the same resolution as our main simulations,
but with a volume $8 \times$ smaller.  While this volume does not
sample massive galaxies well, it does illustrate the main trends.  In
particular, we show the hot gas model because it exhibits a
well-populated red sequence distinct from the blue cloud.  We show marginally resolved galaxies with $r-$band $>-20$ to help clarify the underlying behavior of the simulations.

As described in the text, the merger quenching model under-produces
red galaxies while suppressing overall galaxy growth (compared to a
model without quenching).  This is true for any value of the critical
mass ratio.  Furthermore, the sparsely populated region of red
galaxies is poorly separated from the blue cloud, representing the
tail of a unimodal distribution rather than one peak of a bimodal
distribution.  As we move to higher values of the mass ratio, overall
galaxy growth is suppressed more and more, but only a few galaxies at
any given time are red.  Even with frequent quenching events due to
minor mergers, galaxies continuously accrete new material from the IGM
to fuel star-formation.

\section{Physical Impact of Quenching}
\label{sec.physical}

Our quenching models have many physical and observable implications
for galaxies and the IGM.  Winds from merger quenching could impact
the diffuse IGM.  Our hot halo quenching has a major impact
because it changes star formation histories.  Since outflows
associated with star formation enrich the IGM and often end up
returning to galaxies, changes in the star-formation rates can lead to
changes in IGM and galaxy properties in complicated ways.  Exploring
all these changes is beyond the scope of this work, but here we
explore two physical consequences of quenching -- feedback energetics,
and the overall build-up of the red sequence.

\subsection{Quenching energetics}
\begin{figure*}
\includegraphics[width=165mm]{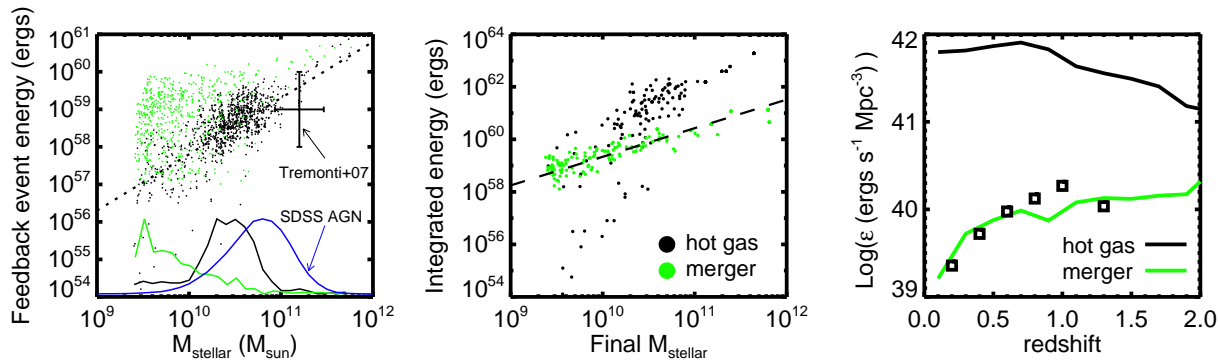}
\caption{Relations between quenching energetics and galaxies, for both
  merger winds (green) and hot gas heating (black).  {\bf Left:}
  Energy associated with each quenching event vs. the stellar mass of
  the galaxy hosting that event.  Since hot gas heating is actually
  continuous, each ``event'' is ambiguous, but corresponds to
  $\sim10^7$ yrs.  Errorbars show an estimate of the energy output in
  winds from post-starburst galaxies based on \citet{tremonti07}.
  Corresponding color-coded histograms of stellar mass are shown at
  the bottom, with blue for optically-selected SDSS AGN.  The
  dotted line shows a scaling $E\propto M^{2/3}$ motivated by the
  hydrostatic temperature scaling with halo mass.  {\bf Middle:}
  Time-integrated energy from quenching feedback as a function of
  $z=0$ galaxy stellar mass.  This is the total energy of feedback
  processes over each galaxy's lifetime, but excludes feedback events
  that occur in small galaxies that merge with the main progenitor.
  The dashed line shows the slope of the
  $M_{\rmn{BH}}-M_{\rmn{stellar}}$ relation for galactic bulges from
  \citet{bennert10}.  {\bf Right:} Total energy density production
  rate of feedback processes vs. redshift.  Squares show a scaled
  estimate of the total luminosity density of AGN from
  \citet{barger05}.}
\label{fig.energetics}
\end{figure*}

In our simple implementations of quenching mechanisms, we have made no
effort to limit the energy available to, e.g., that accreted onto a
supermassive black hole.  We simply eject or heat up the prescribed
gas particles.  Here we investigate whether the energy added via
quenching is reasonable and in line with expectations.


We track the energy associated with our quenching implementations on
galaxy-by-galaxy and event-by-event bases.  For hot halo quenching
event, we add up the change in energy for each gas particle affected,
or quenched, by the event.  With hot gas heating, each heated gas
particle changes in energy by $\Delta E = 3/2 Nk_B \Delta T =
3/2 (M_{\rmn{particle}} / (\mu m_p)) k_B \Delta T$, where
$M_{\rmn{particle}}$ is the total mass of the particle, $\mu=0.6$ is a
typical mean molecular weight for gas in clusters \citep{rosati02},
$m_p$ is the proton mass, $k_B$ is Boltzmann's constant, and $\Delta
T$ is the temperature change of the gas particle.  Since each heated
particle is associated with a galaxy from the group finder, we
attribute its change in energy to the feedback energy tally for that
galaxy.

In our merger quenching model, we estimate the feedback energy for
each particle as the final kinetic energy: $E = 0.5 M_p
v_{\rmn{kick}}^2$.  Here $v_{\rmn{kick}}$ is the kick velocity of gas
ejected in winds associated with mergers.  The total feedback energy
from a single merger is then summed over all particles that
experienced the feedback in a galaxy.  Thus, for our basic model, the
total energy in a merger is effectively $E_{\rmn{merger}}=0.5
M_{\rmn{cold gas}} (1500 \rmn{km s}^{-1})^2$, where $M_{\rmn{cold
    gas}}$ is the total cold gas mass in the merger remnant at the
time of expulsion.

In Figure \ref{fig.energetics}, we compare our quenching energy
calculations to observations associated with feedback processes.  In
the left panel, we show the energy of each feedback event as a
function of its host galaxy's stellar mass.  Since hot gas quenching
is actually continuous, there are too many points to plot clearly, so
we randomly choose 1000 events.  The event energy rises steeply with
stellar mass since more massive halos have more gas that requires
heating to a higher virial temperature.  Since hydrostatic haloes
follow the scaling $M\propto T^{3/2}$, we overplot a dotted line
with $E \propto M_{\rmn{stellar}}^{2/3}$ (with arbitrary
normalization).  The hot gas quenching events follow this trend,
although with some scatter, suggesting that the virial temperature is the
most important factor in determining how much energy is injected in
this model.

Merger quenching shows more scatter and less steep dependence on
stellar mass, both of which are driven by gas fraction trends.  The
event energy for mergers is solely determined by the cold gas mass
(since $v_{\rmn{kick}}$ is the same for all galaxies), so the scatter
in the relation owes to variations in the cold gas mass at any fixed
stellar mass.  In our simulations, more massive galaxies tend to have
smaller gas fractions, which counteracts the overall increase in mass
in these systems, leading to a sub-linear relation between merger
event energy and host galaxy stellar mass.

For comparison with the merger quenching energetics, we estimate the
energy output associated with the massive outflows observed in
post-starburst galaxies.  \citet{tremonti07} measured outflow
velocities $\sim 1000$ km s$^{-1}$, and estimated outflow masses at
$M_{\rmn{wind}}\sim 10^{9-11} M_{\sun}$.  From this we estimate the
kinetic energy as $0.5 M_{\rmn{wind}} v_{\rmn{wind}}^2$, using the
full range of mass estimates to get the errorbars, but neglecting
(substantial) variations in wind velocity.  \citet{tremonti07} also
measured stellar masses for their sample of post-starburst galaxies,
and we plot errorbars based on the standard deviation of the logarithm
of those masses.  Our merger outflow energies fall at the high end of
this observational estimate, entirely consistent given that we used
$v_{\rmn{wind}} = 1500$ km s$^{-1}$ in the simulation.

We also show histograms of the stellar mass distribution of feedback
events along the $x$-axis.  As indicated by the black histogram, most
hot gas quenching events are associated with galaxy stellar masses
between $10^{10}$ and $10^{11} M_{\sun}$.  Because of the steep cutoff
in the galaxy stellar mass function, there are few galaxies above this
mass, and galaxies at lower masses do not have hot halos.  In
contrast, the distribution of stellar masses of merger events (green
line) is skewed toward the low mass end, reflecting the greater
frequency of mergers of low-mass galaxies when averaged over cosmic
time.  For comparison we show in blue the distribution of stellar
masses of AGN host galaxies for AGN selected from SDSS using the
Baldwin-Phillips-Terlevich \citep{BPT} diagnostic with SDSS optical
spectra \citep{kauffmann03_agn_hosts}.  The general trend is similar
though the predicted histogram is shifted to somewhat lower masses.
This indirectly suggests that AGN activity is indicative of ongoing hot halo
quenching.

In the middle panel of Figure \ref{fig.energetics}, we show the
time-integrated feedback energy as a function of host galaxy $z=0$
stellar mass.  If the feedback energy originated from black hole
accretion, this enables comparison with final black hole masses,
modulo corrections for the efficiency at which accreted mass is
converted to output energy, and the coupling of that energy to the
gas.  As with individual event energies, the integrated energy rises
more steeply with stellar mass for hot gas heating than for merger
winds.  A dashed line shows the slope of the
$M_{\rmn{BH}}-M_{\rmn{stellar}}$ relation for local galactic bulges
from \citet{bennert10}, assuming that the integrated feedback energy
is $\propto M_{\rmn{BH}}$.  The amplitude is arbitrarily normalized to
coincide with the mean energy for merger winds at the low-mass end.
The slope for merger winds, though slightly shallower than the
$M_{\rmn{BH}}-M_{\rmn{stellar}}$ relation, is a good match to the
observations.  The slope for hot gas quenching is also in good
agreement at the high mass end, but drops away quickly to low masses
where little energy input has occured.  Note that the observed
$M_{\rmn{BH}}-M_{\rmn{stellar}}$ relation uses the {\it bulge} stellar
mass, whereas the simulated galaxies show the total stellar mass.
Hence the dropoff to low masses in the hot gas quenching case may
partly be reconciled by lower bulge-to-disk ratios in sub-$L^\star$
galaxies.

Finally, the right plot shows the energy density production rate of
quenching as a function of redshift.  This is analogous to the plot of
\citet{madau98} of galaxy luminosity density evolution, and directly
comparable to the plot of \citet{barger05} of AGN luminosity density
evolution \citep[see also][]{cowie04}.  We take data from Figure 25 of
\citeauthor{barger05} for spectroscopically identified sources with
$L_{2-10\rmn{ keV}}\ga 10^{42}$ ergs s$^{-1}$ and multiply by their
suggested bolometric correction of 35 to get the total luminosity.
Then we multiply by a coupling efficiency which determines what
fraction of an AGN's total luminosity couples to the surrounding gas.
We choose a somewhat high value of 0.1 \citep[0.05 is
typical][]{dimatteo05} to match the normalization of the energy
density production rate for the merger quenching model, and show the
result as squares in the figure.  Outflows from galaxy mergers in our
simulations match the shape of measured energetics of luminous AGNs.
The total energy requirements for hot gas heating appear much higher.
This may indicate that hot gas quenching in the real universe (1) is
tied to lower-luminosity AGN, (2) does not heat gas up to the virial
temperature, and/or (3) is intermittent with a small duty cycle.


Overall, our results are broadly consistent with the idea that AGN are
responsible for providing the energy for quenching feedback.  Various
authors have already shown that merger-driven feedback can lead to
galaxies that are consistent with the $M-\sigma$ relation; our
simulations likewise appear to do so.  However, our simulations also
show that merger quenching is, in itself, unlikely to be responsible
for producing red and dead galaxies.  The energetic requirements of
hot gas quenching are quite severe in our current model
implementation. For instance, such energy injection will excessively 
heat intragroup gas, likely leading to a mismatch with observed temperature and entropy profiles; we plan to explore this in future work.
We speculate that intermittent heating, perhaps triggered by
an accumulation of cold gas in the halo center, might work just as
well to quench star formation with a significantly lower energy budget.
Hence while forcing gas in hot halos to remain at the virial 
temperature produces a reasonable red galaxy population, further work
is required to construct a fully viable physical model.

\subsection{Red sequence build-up over cosmic time}

\begin{figure}
\includegraphics[width=84mm]{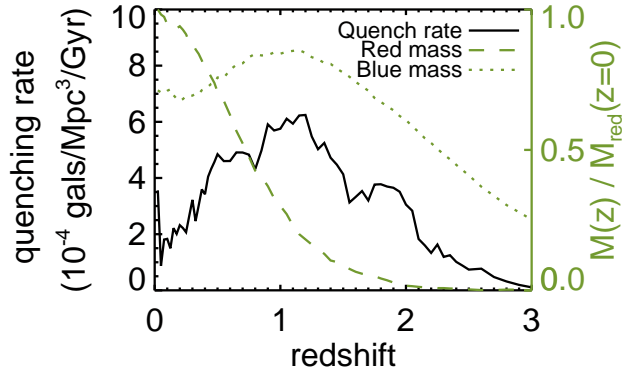}
\caption{Quenching rate (black, left axis) and mass growth history
  normalized to $z=0$ (green, right axis) of red sequence galaxies
  with $M_{\rmn{stellar}} > 10^{9.5} M_{\sun}$ in the hot gas
  quenching model.  In this model, quenching peaks at $z\approx1$,
  declining at late times.  About 2/3 of the mass of the red sequence
  builds up at $z<1$, while the mass in the blue cloud declines during
  the same period.}
\label{fig.quenchrate}
\end{figure}

The build up of the red sequence over time can place strong
constraints on quenching physics.  Here we investigate two simple
parameterizations: the galaxy number quenching rate and the mass
growth histories of the red sequence.  For both quantities, we
restrict ourselves to red galaxies with stellar masses above $10^{9.5}
M_{\sun}$.

We calculate the quenching rate by comparing each consecutive pair of
snapshot outputs from our basic hot gas quenching simulation.  The
quenching rate, $dN_{\rmn{quenched}}/dt$ is estimated as $\Delta
N_{\rmn{low ssfr}} / \Delta t_{\rmn{snap}}$.  Here, $\Delta
N_{\rmn{low ssfr}}$ is the difference in the number of galaxies with
specific star formation rates (SSFRs) below $10^{-12}$ yr$^{-1}$
between the two snapshots, and $\Delta t_{\rmn{snap}}$ is the time
between snapshots.  The red sequence mass growth is the total stellar
mass in such low-SSFR galaxies, $M_{\rm red}(z)$, normalized to the
total stellar mass at in these galaxies $z=0$, $M_{\rm red}(z=0)$.  We
also calculate the mass growth of galaxies with SSFRs above this
cutoff, $M_{\rm blue}(z)$, normalizing to the mass in red galaxies
$M_{\rm red}(z=0)$ to track relative changes in the blue and red
populations.

Figure \ref{fig.quenchrate} shows the results.  The quenching rate is
small at $z>2$, after which it rapidly increases.  The rate peaks at
$z\approx 1$, declining at low redshift.  Observations tend to
indicate that the mean formation time of stellar populations in
massive ellipticals is at $z\ga 2$ \citep[e.g. ][]{graves09a}.  In our model, while
massive galaxies tend to have old stellar populations, they do not
quench their star formation until typically $z\sim 1$.  This may be
related to why the color-density relation disappears or inverts at
$z\ga 1$: At low-$z$, typical central massive galaxies are red,
leading to the familiar positive color-density correlation, while
prior to $z\sim 1$ central galaxies have typically not quenched and
are blue, leading to an inverted relation.

The total mass on the red sequence at $z=1$ is about 1/3 the final
value at $z=0$; half the mass on the present-day red sequence is in
place at $z=0.8$.  This significant late build-up is slightly more
than the factor of 2 in stellar mass \citep{bell04} and galaxy number
\citep{faber07} suggested by observations, although such estimates are
presented as lower limits.  The mass on the blue cloud rises at early
times, declining at $z<1$ as many star-forming galaxies become
quenched.  This reflects a decrease in the number density of blue
galaxies by $\sim 20$\%, substantially different from the empirical
analysis of \citet{blanton06}, which found no evolution in number
density of the observed blue cloud to within 10\%.  By $z=0$, the
total stellar mass in blue galaxies (above our resolution limit of
$\sim 3 \times 10^9 M_{\sun}$) is $\sim 70$\% that in red galaxies.

\section{Discussion}
\label{sec.discussion}


\subsection{Red galaxy morphologies}

In this work we have focused on the transformation of galaxy colours,
but of course colour and morphology are closely correlated in the
local universe.  Our simulations' resolution is poorly suited to study
morphologies~\citep[see e.g.][for a fuller discussion of why]{brook10}.  In our favoured hot gas model, starvation by the hot
halo only stops star-formation without changing the morphology.  Once
star-formation stops, a galaxy must undergo (possibly minor) mergers
to disrupt the remaining disk and form an elliptical.  Such dry
(gas-poor) mergers may have difficulty reproducing the observed
phase-space densities of stellar orbits in the centres of some elliptical galaxies \citep{hopkins09_cusps}, but
some fraction of the bulge may have built up via dissipational mergers
before the galaxy starved \citep[e.g.][]{bournaud07}.

One key argument in favor of gas-rich major mergers as a transformative
process is that they might turn a star-forming disk into a quiescent
elliptical all at once.  However, recent observations of quenched
late-type galaxies disfavor the merger-only scenario, and suggest that
a substantial fraction of passive galaxies must have gone through a
``passive disk'' phase \citep{bundy10, vanderwel10}.  Our results
reinforce the theoretical difficulty seen in earlier models: after
halting star-formation at any given time in a galaxy's evolution,
something must prevent future gas accretion to keep that galaxy red and
dead \citep{hopkins08_ellipticals, croton06, bower06}.

Finally, we note that our hot gas quenching model does not produce
passive field galaxies in low-mass haloes (which are not dominated
by hot gas), and this may violate observational constraints on the
relationship between stellar-mass and the fraction of passive galaxies
\citep{hopkins08_ellipticals}.  Ultimately, morphologies provide an
interesting constraint on quenching models, but higher-resolution
simulations or sub-resolution morphological models are required
to properly apply them.  Insofar as galaxy structure influences
star-formation histories \citep{martig09}, such models may be necessary
to understand red galaxy formation.

\subsection{The importance of hot halo gas}

Our favoured model relies on the hot gas surrounding massive
galaxies.  This hot corona is a robust prediction of years of theoretical
work, yet its existence has not been universally established at mass
scales below groups.  X-ray observations have revealed diffuse hot gas
around only some of the most massive field galaxies \citep{mulchaey10},
but recent simulations predict that the halo gas around $L_{\rm
optical}^*$ galaxies should have X-ray luminosities too faint for current
observatories \citep{crain10}.  The gas may be there, we just cannot
see it.

Why should the hot gas be implicated in shutting down star formation?
The hot gas may be crucial to triggering the \emph{type} of feedback that
best couples with surrounding gas \citep[i.e. radio jets][]{dimatteo00},
or it may simply be a more effective ``net'' than cold IGM to ``catch''
the energetic output of an AGN.  It seems an unlikely coincidence that
the mass scale where a stable hot halo can form just happens to be
around the turnover in the stellar mass function, but observations do
not unambiguously favor a direct connection.  Among field galaxies with
diffuse X-ray gas detections, both star-forming and passive galaxies
are well-represented \citep{crain11}.  This result is not entirely
inconsistent with our model -- after new accretion is quenched, a
galaxy will remain star-forming for $\sim 2$ Gyrs while surrounded by
a substantial hot gaseous halo.  Nevertheless, the underabundance of
massive star-forming galaxies in our favoured model suggests a less direct
relationship between hot gas and quenching than our simple prescription.
Perhaps quenching depends jointly on hot gas and the mass of the central
black hole, which itself is the product of a complicated history that
may involve galaxy mergers and secular growth.

\subsection{Comparison to other models}

Our on-the-fly quenching models build on the post-processing
models of \citet{gabor10}, and the results show some similarities.
The color of the red sequence is too blue and roughly constant with
mass, which we attribute to difficulties with supernova metal yields,
the redistribution of metals by star-formation-driven winds, and/or a
poor treatment of merger-induced starbursts in our simulations.  We also
observe a similar sharp cutoff in the blue galaxy mass function in our
best model.  A major difference, however, is that re-fueling of merger
remnants prevents our merger quenching model from yielding a red sequence,
whereas in \citet{gabor10} we assumed that all future star-formation
was quenched after a merger.

With many successes at matching observations, SAMs also serve as a useful
benchmark for our models.  Unlike our simulations, typical SAMs already
resulted in a substantial red sequence even before invoking quenching
feedback models \citep{croton06, bower06}.  The red galaxies arise due
to rapid gas consumption in a starburst associated with a merger or
disk instability, or due to gas stripping as a satellite galaxy falls
into a larger halo.  Our simulations do produce quenched satellites
\citep{dave11}, but probably do not track secular disk evolution properly.
Nevertheless, our merger quenching model shares the generic feature with
modern SAMs that further ongiong quenching is required to ensure that the
most massive galaxies remain red and dead.  Different models incorporate
different mechanisms for this added quenching, from various radio mode
feedback prescriptions to a simple halo mass cutoff \citep{croton06,
bower06, cattaneo06, somerville08}.  It is worth recognizing that our
late-time re-accretion of gas is dominated by recycled winds, something
that is not dynamically followed in the SAMs.

Cosmological hydrodynamic simulations akin to ours have included various
feedback processes, but (to our knowledge) none has previously produced
a realistic red sequence and mass/luminosity function.  Simulations with
neither stellar nor black hole feedback produce massive galaxies with
low star-formation rates, but their stellar masses are much too large
\citep{keres09_feedback}.  Our simulations with only stellar feedback
can bring the low-mass end of the galactic stellar mass function into
agreement with observations, but massive galaxies form stars at excessive
rates \citep{oppenheimer10}.  By coupling sub-grid Bondi-Hoyle-Lyttleton
black hole accretion with models for the associated energy output, other
simulations show promise for matching black hole growth and suppressing
star-formation in massive galaxies \citep{sijacki07, dimatteo08, booth09,
booth10, degraf10, mccarthy10}, but it is unclear whether they produce a
red sequence as observed.  Understanding the differences between these
models and ours will provide complementary insights into the physical
processes behind quenching.

\subsection{Star-formation driven winds and quenching}
\label{sec.winds}

\citet{oppenheimer10} highlighted the importance of wind recycling
accretion in our simulations (without quenching), showing that
star-formation in galaxies with
$M_{\rmn{stellar}}\gtrsim10^{10}M_{\sun}$ is fueled mainly by gas that
was earlier ejected in winds.  Because this wind recycling dominates
the fuel for star-formation in massive galaxies, it is the main
fueling mode we are trying to quench in this work.

Our treatment of star-formation driven winds may not be physically
accurate.  Once launched, we decouple the winds from hydrodynamic
interactions with surrounding gas to allow them to escape the galaxy
in a resolution-converged way \citep{springel03, schaye10}.  Even
after re-activation of the hydrodynamics, the wind particles may not
merge seamlessly with surrounding gas.  Upon escape from galaxies,
wind particles are cold, dense, and metal-rich.  While they decrease
in density as they move away from the galaxy, they remain enriched
since we don't include diffusion of metals among gas particles.
Metal-line cooling makes them cool faster than surrounding gas, so
they stay cold and dense, and more likely to return to galaxies.  It
is unclear how realistic the assumed complete lack of mixing is, but
we note that this wind model has resulted in agreement with
observations of IGM metal absorbers
\citep{oppenheimer06,oppenheimer08,oppenheimer09_metal_absorption}.

Wind recycling, and the possibility of over-recycling, has
implications for all our quenching models.  As shown in
\citet{oppenheimer10} and \citet{gabor10}, simply stopping wind
recycling is unlikely to produce a realistic red sequence -- doing so
suppresses star formation too much to create enough $\sim L^*$
galaxies, while small amounts of cold and hot accretion keep too many
galaxies forming stars.  In our merger quenching model, we have
emphasized the re-fueling of merger remnants after $\sim 2$ Gyr.
Based on the origin of star-forming gas explored in
\citet{oppenheimer10}, most of the accreted fuel comes from wind
recycling, although a significant sub-dominant component comes from
cold and hot accretion.  Therefore in our hot gas quenching model, where we
expressly heat all non-star-forming gas, most of the gas we must
quench had been expelled in winds.  Hence we caution that the
hydrodynamics of winds may play a critical role in quantitatively
assessing the amount and location of quenching feedback needed to
produce a red sequence.


\subsection{Guidelines for future models}


Given the wide range of physical scales important in galaxy
evolution, developers of cosmological simulations have little choice but
to resort to sub-resolution quenching models for the foreseeable future.
Here we discuss ways of improving these models. Any model quenching
mechanism requires 3 steps:

\begin{enumerate}
\item Criteria to start (and stop) the feedback process.
\item An estimate of the energy budget or luminosity available.
\item A method to couple feedback energy to surrounding gas.
\end{enumerate}

The first two steps may be linked in models where the feedback
energy varies over time: e.g. the black hole accretion rate varies
by several orders of magnitude between states where the feedback is
important and negligible. In our current hot gas quenching model,
(i) is the existence of a hot gas-dominated halo, (ii) is effectively
unlimited, and for (iii) we heat all the non-star-forming gas in the {\sc
FOF} group.  We have seen that this successfully creates a red sequence
of galaxies, but that it leads to a dearth of star-forming galaxies
with $M_{\rmn{stellar}}\gtrsim10^{10.5}M_{\sun}$ and requires too
much energy input.  Below we discuss each of these steps in the context
of radio mode AGN feedback.


\subsubsection{(i) Triggering mechanisms}

Triggering radio mode feedback requires accretion onto a supermassive
black hole.  In massive (cluster) haloes, the black hole may accrete
directly from the hot halo in a Bondi process \citep{quataert00}, or a
black hole may be fueled by cold clouds that condense from the hot
medium \citep{soker06, mcnamara10}.  Bondi accretion of hot gas may be
more conducive to the formation of a powerful radio jet than rapid,
radiatively efficient accretion \citep{dimatteo00}.  The formation of
a jet may additionally require significant black hole spin
\citep{narayan05}.  Based on the present work, the presence of hot gas
seems to be crucial for triggering radio mode feedback.

Models have used a variety of triggers for AGN feedback.  SAMs
typically assume that feedback operates in hot, massive haloes.
Hydrodynamic models usually adopt a Bondi-Hoyle prescription, or a
simple scaling with gas mass in the black hole vicinity \citep{vernaleo06, bruggen09}, to control
black hole accretion and thus triggering.

Observationally, most X-ray bright clusters, groups, and elliptical
galaxies host some level of central radio AGN, with AGN more likely in
systems with shorter central cooling times, and more luminous in more
massive systems \citep{burns90, mittal09, sun09, dunn10}.  Outbursts
that induce X-ray cavities typically occur every $10^{7-8}$ years
with energies $\sim 10^{55-60}$ ergs \citep{voit05,
david10,randall10}.  These observations suggest a triggering mechanism
linked to gas cooling in the central regions of a hot halo, which may
be a resolvable triggering mechanism in cosmological simulations.
This may provide a more observationally-calibratable method for
triggering AGN feedback than Bondi-Hoyle accretion.

\subsubsection{(ii) Available energy for feedback}

The energy of feedback mainly depends on the black hole accretion
rate.  Most models assume some fraction of the accretion rest-mass-energy
acts on surrounding gas.  The accretion rate is typically determined
through simple scalings with black hole and halo mass in SAMs and
through Bondi-Hoyle models in simulations.


Observations suggest kinetic power output from central bursts of $\sim
10^{45}$ erg s$^{-1}$ over $\sim 10^8$ years in some clusters,
consistent with total energy in inflated bubbles ($\sim 10^{60}$
ergs).  As mentioned above, radio emission scales with cluster size,
but it is not clear to what extent the burst energies depend on halo
properties.  The fact that few very massive galaxies have enough
star-formation to give them blue colors suggests that there is enough
heating power to offset cooling.  This may imply a self-regulating
cycle \citep[cf.][]{ciotti97, ciotti01, churazov05}.  Hydrodynamic
models that directly track black hole accretion or use some estimate
based on the mass within the central regions of the halo can achieve
this self-regulation \citep[e.g.][]{bruggen09}.

In our hot gas quenching model, the required energy budget is large,
although it is not clear that significantly less energy input could be
equally effective.  It is also difficult to exceed the theoretical
maximum energy output of black holes; models typically assume
efficiency of mass-energy coupling to surrounding gas of less than
1\%.  So the energy budget remains an interesting constraint, but the
most attractive scenario is that black holes emit just enough energy
to remove or starve their cold gas reservoir, at which point they
become quiescent again.

\subsubsection{(iii) Coupling of feedback to gas}

How AGN energy couples with the surrounding medium is another matter
of debate.  Supersonic jets induce bow shocks in the surrounding gas
\citep{voit05, randall10}, and subsonic jets entrain material and
thermalize their energy via turbulent mixing \citep{deyoung10}.  The
outflows can apparently evacuate $\sim 10$ kpc-scale ``bubbles'' in hot
intragroup gas \citep{birzan04}, and the bubbles then share their
energy at scales $\sim 10-100$ kpc via buoyant uplift, mixing, and
sound waves \citep{churazov01, voit05, nulsen07, scannapieco08, bruggen09,fabian03, sanders07,sanders08}.

While idealized hydrodynamic models have directly examined the
interaction of jets and bubbles with the ICM \citep[e.g.][]{ruszkowski04b, bruggen05, heinz06, scannapieco08, bruggen09_bubbles, morsony10, mendygral11}, SAMs typically assume
that feedback directly offsets the cooling for simplicity.
Cosmological hydrodynamic simulations usually inject the feedback
energy into one or a few resolution elements near the black hole,
relying on the hydrodynamic calculations to distribute the energy to
large scales in the halo \citep[e.g.][]{booth09, teyssier11}.  In our models, suppressing star-formation
in accordance with observations requires heating gas throughout the
circum-galactic region, on scales of $>10$ kpc, to prevent the
formation of cold clumps.

In summary, a path for modeling quenching feedback 
would be as follows: 1) Trigger
feedback when the central cooling time drops below e.g. 1 Gyr, or some
fraction of the hubble time; 2) Calculate energy for feedback as some
fraction of the rest-mass energy of cold or cooling gas in the central
regions; 3) Increase the entropy of the gas with some radial
dependence in a manner consistent with higher-resolution simulations.
We will explore models based on these principles in future work.

\section{Summary and Conclusion} \label{sec.conclusion} 

With the goal of building a realistic red sequence of galaxies in
cosmological hydrodynamic simulations, we have implemented novel
mechanisms for quenching star-formation in the simulation code
GADGET-2.  By identifying mergers and halos on-the-fly within
simulations, we implement and test various models for quenching
feedback related to these processes.  While our work is motivated by
feedback resulting from AGN, we explicitly avoid examining black hole
growth in order to concentrate on constraints from the massive galaxy
population.

Motivated by observations of massive outflows from post-starburst
galaxies attributed to quasar feedback, we implement $1500$ km
s$^{-1}$ superwinds in the remnants of galaxy mergers that expels all
the cold gas.  To do so, we first use an {\sc fof} group finder to
identify galaxy mergers on-the-fly, and then give a velocity kick to
all the star-forming gas.  We have shown that this quenching mechanism
alone does not produce a red sequence in our simulations.  Even after
all the gas is expelled from a merger remnant, new supplies of gas
accrete from the IGM to re-fuel star-formation, typically within $\sim
2$ Gyr.

Motivated by observations of radio AGN and X-ray cavities in the hot
gas of galaxy groups, we add thermal heating to hot gas in massive
dark matter haloes.  We calculate the hot gas fraction $f_{\rmn{hot}}$
in the halo, and if $f_{\rmn{hot}}>0.6$ (roughly corresponding to halo
masses $\gtrsim 10^{12}M_{\sun}$), then we heat all its gas outside of
the ISM to the halo virial temperature.  By keeping the surrounding
gas hot, we starve galaxies embedded in hot haloes of new fuel for
star-formation.  We have shown that this hot gas quenching model
yields a red sequence whose luminosity function provides an excellent
match with observations of local galaxies (Figure
\ref{fig.cmdlf_main}).

Our main results are:
\begin{itemize}
\item Galactic-scale outflows triggered by mergers (i.e. quasar mode
  feedback) do not produce a substantial red sequence on their own
  because gas accretion from the IGM re-fuels star-formation within
  $\sim1-2$ Gyr.
\item Adding thermal energy to hot X-ray gas around massive galaxies
  (analogous to radio mode AGN feedback) successfully produces a red
  sequence whose luminosity function matches observations.
\item This heating must occur around satellite galaxies embedded in
  the hot gas of their parent haloes to match the faint-end red galaxy
  luminosity function.
\item A combination of hot gas heating with merger-triggered outflows
  may be empirically motivated and perform as well as the heating-only
  model, but hot gas heating is the crucial required component.
\item Our simple hot gas heating model produces somewhat too few
  massive blue galaxies, possibly owing to the sharp truncation in gas
  accretion onto galaxies as soon as their haloes are dominated by hot
  gas.
\item As in \citet{gabor10}, our baseline model produces a red sequence
  that is too blue and too shallow, likely owing to issues related to
  enrichment.  We empirically recalibrate the metallicities to obtain
  agreement, which requires a metallicity increase up to $\times2$ in
  the most massive systems.
\end{itemize}

The overall success of reproducing the observed red sequence and
associated luminosity function is a first for cosmological hydrodynamic
simulations.  Doing so has already elucidated stringent constraints on
how quenching must operate under the scenarios explored.  While hot
halo quenching appears to be necessary and sufficient to reproduce
observations of red galaxies as well as any model at $z=0$,
our current simplistic implementation requires more energy than is
thought to be available and likely overheats surrounding gas. We have
yet to explore details of redshift evolution, clustering, or the impact
of quenching on the surrounding intergalactic gas; these may motivate
variants on this simple scenario.  The beauty of implementing quenching
models self-consistently within large-scale hydrodynamics simulations
of galaxy formation is that such models open up a host of new avenues
to constrain quenching physics.

In future work, we aim to develop more physically consistent models
for quenching star-formation.  To do so, we advocate connecting
well-resolved structures and processes in the simulations to feedback
processes that operate below the resolution scale.  Such
sub-resolution models will require guidance from observations and
higher-resolution simulations of individual galaxies or clusters.  We
outline a particular path forward for this based on current intuition
that we will explore in future work.  By combining insights from
advancing theoretical work on all scales and advancing observations
across cosmic time, we hope to continue refining our models to better
understand how massive red and dead galaxies come to be.


 
 

\section*{Acknowledgments}
The authors acknowledge N. Katz, D. Keres, J. Kollmeier, A. van der
Wel, and D. Weinberg for helpful discussions, the referee for useful
comments, and V. Springel for making Gadget-2 public. The simulations
used here were run on University of Arizona's SGI cluster, ice. This
work was supported by the National Science Foundation under grant
numbers AST-0847667 and AST- 0907998. Computing resources were
obtained through grant number DMS-0619881 from the National Science
Foundation.  KF acknowledges support from NASA through Hubble
Fellowship grant HF-51254.01 awarded by the Space Telescope Science
Institute, which is operated by the Association of Universities for
Research in Astronomy, Inc., for NASA, under contract NAS 5-26555.

\bibliographystyle{mn2e} 

\bibliography{paper}


\clearpage

\label{lastpage}
\end{document}